\documentclass[longauth,twocolumn,traditabstract]{aa}  
\usepackage{fixltx2e}
\usepackage[english]{babel}
\usepackage{graphicx}
\usepackage{epsf,color}
\usepackage[mathscr]{eucal}
\usepackage[fleqn]{amsmath}
\usepackage{amssymb,amsfonts}
\usepackage{natbib}
\usepackage{txfonts}
\usepackage{dsfont}
\usepackage{color}
\usepackage{floatrow}
\usepackage{subfig}
\usepackage{caption}
\usepackage{widetext}
\usepackage{Euclid}
\bibpunct{(}{)}{;}{a}{}{,}
\begin{document}
\title{\Euclid: Estimation of the impact of correlated readout noise for flux measurements with the \Euclid NISP instrument\thanks{This paper is published on behalf of the Euclid Consortium.}}

\author{A.~Jim\'enez Mu\~noz$^{1}$\thanks{\email{jimenez@lpsc.in2p3.fr}}, J.~Mac\'{\i}as-P{\'e}rez$^{1}$, A.~Secroun$^{2}$, W.~Gillard$^{2}$, B.~Kubik$^{3}$, N.~Auricchio$^{4}$, A.~Balestra$^{5}$, C.~Bodendorf$^{6}$, D.~Bonino$^{7}$, E.~Branchini$^{8,9,10}$, M.~Brescia$^{11}$, J.~Brinchmann$^{12,13}$, V.~Capobianco$^{7}$, C.~Carbone$^{14}$, J.~Carretero$^{15}$, R.~Casas$^{16,17}$, M.~Castellano$^{10}$, S.~Cavuoti$^{11,18,19}$, A.~Cimatti$^{20,21}$, R.~Cledassou$^{22,23}$, G.~Congedo$^{24}$, L.~Conversi$^{25,26}$, Y.~Copin$^{27}$, L.~Corcione$^{7}$, A.~Costille$^{28}$, M.~Cropper$^{29}$, H.~Degaudenzi$^{30}$, M.~Douspis$^{31}$, F.~Dubath$^{30}$, S.~Dusini$^{32}$, A.~Ealet$^{27}$, E.~Franceschi$^{4}$, P.~Franzetti$^{14}$, M.~Fumana$^{14}$, B.~Garilli$^{14}$, B.~Gillis$^{24}$, C.~Giocoli$^{33,34}$, A.~Grazian$^{5}$, F.~Grupp$^{6,35}$, S.V.H.~Haugan$^{36}$, W.~Holmes$^{37}$, F.~Hormuth$^{38,39}$, K.~Jahnke$^{39}$, S.~Kermiche$^{2}$, A.~Kiessling$^{37}$, M.~Kilbinger$^{40}$, M.~K\"ummel$^{35}$, M.~Kunz$^{41}$, H.~Kurki-Suonio$^{42}$, R.~Laureijs$^{43}$, S.~Ligori$^{7}$, P.~B.~Lilje$^{36}$, I.~Lloro$^{44}$, E.~Maiorano$^{4}$, O.~Mansutti$^{45}$, O.~Marggraf$^{46}$, K.~Markovic$^{37}$, R.~Massey$^{47}$, E.~Medinaceli$^{33}$, S.~Mei$^{48}$, M.~Meneghetti$^{4,49,50}$, G.~Meylan$^{51}$, L.~Moscardini$^{4,20,52}$, S.M.~Niemi$^{43}$, C.~Padilla$^{15}$, S.~Paltani$^{30}$, F.~Pasian$^{45}$, K.~Pedersen$^{53}$, W.J.~Percival$^{54,55,56}$, S.~Pires$^{40}$, G.~Polenta$^{57}$, M.~Poncet$^{23}$, L.~Popa$^{58}$, L.~Pozzetti$^{4}$, F.~Raison$^{6}$, R.~Rebolo$^{59,60}$, M.~Roncarelli$^{20,33}$, E.~Rossetti$^{20}$, R.~Saglia$^{6,35}$, M.~Sauvage$^{40}$, R.~Scaramella$^{10,61}$, P.~Schneider$^{46}$, G.~Seidel$^{39}$, S.~Serrano$^{16,17}$, C.~Sirignano$^{32,62}$, G.~Sirri$^{50}$, D.~Tavagnacco$^{45}$, A.N.~Taylor$^{24}$, H.I.~Teplitz$^{63}$, I.~Tereno$^{64,65}$, R.~Toledo-Moreo$^{66}$, L.~Valenziano$^{4,50}$, T.~Vassallo$^{35}$, G.A.~Verdoes Kleijn$^{67}$, Y.~Wang$^{63}$, J.~Weller$^{6,35}$, M.~Wetzstein$^{6}$, G.~Zamorani$^{4}$, J.~Zoubian$^{2}$}

\institute{$^{1}$ Univ. Grenoble Alpes, CNRS, Grenoble INP, LPSC-IN2P3, 53, Avenue des Martyrs, 38000, Grenoble, France\\
$^{2}$ Aix-Marseille Univ, CNRS/IN2P3, CPPM, Marseille, France\\
$^{3}$ University of Lyon, UCB Lyon 1, CNRS/IN2P3, IUF, IP2I Lyon, France\\
$^{4}$ INAF-Osservatorio di Astrofisica e Scienza dello Spazio di Bologna, Via Piero Gobetti 93/3, I-40129 Bologna, Italy\\
$^{5}$ INAF-Osservatorio Astronomico di Padova, Via dell'Osservatorio 5, I-35122 Padova, Italy\\
$^{6}$ Max Planck Institute for Extraterrestrial Physics, Giessenbachstr. 1, D-85748 Garching, Germany\\
$^{7}$ INAF-Osservatorio Astrofisico di Torino, Via Osservatorio 20, I-10025 Pino Torinese (TO), Italy\\
$^{8}$ INFN-Sezione di Roma Tre, Via della Vasca Navale 84, I-00146, Roma, Italy\\
$^{9}$ Department of Mathematics and Physics, Roma Tre University, Via della Vasca Navale 84, I-00146 Rome, Italy\\
$^{10}$ INAF-Osservatorio Astronomico di Roma, Via Frascati 33, I-00078 Monteporzio Catone, Italy\\
$^{11}$ INAF-Osservatorio Astronomico di Capodimonte, Via Moiariello 16, I-80131 Napoli, Italy\\
$^{12}$ Centro de Astrof\'{\i}sica da Universidade do Porto, Rua das Estrelas, 4150-762 Porto, Portugal\\
$^{13}$ Instituto de Astrof\'isica e Ci\^encias do Espa\c{c}o, Universidade do Porto, CAUP, Rua das Estrelas, PT4150-762 Porto, Portugal\\
$^{14}$ INAF-IASF Milano, Via Alfonso Corti 12, I-20133 Milano, Italy\\
$^{15}$ Institut de F\'{i}sica d’Altes Energies (IFAE), The Barcelona Institute of Science and Technology, Campus UAB, 08193 Bellaterra (Barcelona), Spain\\
$^{16}$ Institute of Space Sciences (ICE, CSIC), Campus UAB, Carrer de Can Magrans, s/n, 08193 Barcelona, Spain\\
$^{17}$ Institut d’Estudis Espacials de Catalunya (IEEC), Carrer Gran Capit\'a 2-4, 08034 Barcelona, Spain\\
$^{18}$ Department of Physics "E. Pancini", University Federico II, Via Cinthia 6, I-80126, Napoli, Italy\\
$^{19}$ INFN section of Naples, Via Cinthia 6, I-80126, Napoli, Italy\\
$^{20}$ Dipartimento di Fisica e Astronomia “Augusto Righi” - Alma Mater Studiorum Università di Bologna, via Piero Gobetti 93/2, I-40129 Bologna, Italy\\
$^{21}$ INAF-Osservatorio Astrofisico di Arcetri, Largo E. Fermi 5, I-50125, Firenze, Italy\\
$^{22}$ Institut national de physique nucl\'eaire et de physique des particules, 3 rue Michel-Ange, 75794 Paris C\'edex 16, France\\
$^{23}$ Centre National d'Etudes Spatiales, Toulouse, France\\
$^{24}$ Institute for Astronomy, University of Edinburgh, Royal Observatory, Blackford Hill, Edinburgh EH9 3HJ, UK\\
$^{25}$ European Space Agency/ESRIN, Largo Galileo Galilei 1, 00044 Frascati, Roma, Italy\\
$^{26}$ ESAC/ESA, Camino Bajo del Castillo, s/n., Urb. Villafranca del Castillo, 28692 Villanueva de la Ca\~nada, Madrid, Spain\\
$^{27}$ Univ Lyon, Univ Claude Bernard Lyon 1, CNRS/IN2P3, IP2I Lyon, UMR 5822, F-69622, Villeurbanne, France\\
$^{28}$ Aix-Marseille Univ, CNRS, CNES, LAM, Marseille, France\\
$^{29}$ Mullard Space Science Laboratory, University College London, Holmbury St Mary, Dorking, Surrey RH5 6NT, UK\\
$^{30}$ Department of Astronomy, University of Geneva, ch. d\'Ecogia 16, CH-1290 Versoix, Switzerland\\
$^{31}$ Universit\'e Paris-Saclay, CNRS, Institut d'astrophysique spatiale, 91405, Orsay, France\\
$^{32}$ INFN-Padova, Via Marzolo 8, I-35131 Padova, Italy\\
$^{33}$ Istituto Nazionale di Astrofisica (INAF) - Osservatorio di Astrofisica e Scienza dello Spazio (OAS), Via Gobetti 93/3, I-40127 Bologna, Italy\\
$^{34}$ Istituto Nazionale di Fisica Nucleare, Sezione di Bologna, Via Irnerio 46, I-40126 Bologna, Italy\\
$^{35}$ Universit\"ats-Sternwarte M\"unchen, Fakult\"at f\"ur Physik, Ludwig-Maximilians-Universit\"at M\"unchen, Scheinerstrasse 1, 81679 M\"unchen, Germany\\
$^{36}$ Institute of Theoretical Astrophysics, University of Oslo, P.O. Box 1029 Blindern, N-0315 Oslo, Norway\\
$^{37}$ Jet Propulsion Laboratory, California Institute of Technology, 4800 Oak Grove Drive, Pasadena, CA, 91109, USA\\
$^{38}$ von Hoerner \& Sulger GmbH, Schlo{\ss}Platz 8, D-68723 Schwetzingen, Germany\\
$^{39}$ Max-Planck-Institut f\"ur Astronomie, K\"onigstuhl 17, D-69117 Heidelberg, Germany\\
$^{40}$ AIM, CEA, CNRS, Universit\'{e} Paris-Saclay, Universit\'{e} de Paris, F-91191 Gif-sur-Yvette, France\\
$^{41}$ Universit\'e de Gen\`eve, D\'epartement de Physique Th\'eorique and Centre for Astroparticle Physics, 24 quai Ernest-Ansermet, CH-1211 Gen\`eve 4, Switzerland\\
$^{42}$ Department of Physics and Helsinki Institute of Physics, Gustaf H\"allstr\"omin katu 2, 00014 University of Helsinki, Finland\\
$^{43}$ European Space Agency/ESTEC, Keplerlaan 1, 2201 AZ Noordwijk, The Netherlands\\
$^{44}$ NOVA optical infrared instrumentation group at ASTRON, Oude Hoogeveensedijk 4, 7991PD, Dwingeloo, The Netherlands\\
$^{45}$ INAF-Osservatorio Astronomico di Trieste, Via G. B. Tiepolo 11, I-34131 Trieste, Italy\\
$^{46}$ Argelander-Institut f\"ur Astronomie, Universit\"at Bonn, Auf dem H\"ugel 71, 53121 Bonn, Germany\\
$^{47}$ Institute for Computational Cosmology, Department of Physics, Durham University, South Road, Durham, DH1 3LE, UK\\
$^{48}$ APC, AstroParticule et Cosmologie, Universit\'e Paris Diderot, CNRS/IN2P3, CEA/lrfu, Observatoire de Paris, Sorbonne Paris Cit\'e, 10 rue Alice Domon et L\'eonie Duquet, 75205, Paris Cedex 13, France\\
$^{49}$ California institute of Technology, 1200 E California Blvd, Pasadena, CA 91125, USA\\
$^{50}$ INFN-Sezione di Bologna, Viale Berti Pichat 6/2, I-40127 Bologna, Italy\\
$^{51}$ Observatoire de Sauverny, Ecole Polytechnique F\'ed\'erale de Lau- sanne, CH-1290 Versoix, Switzerland\\
$^{52}$ INFN-Bologna, Via Irnerio 46, I-40126 Bologna, Italy\\
$^{53}$ Department of Physics and Astronomy, University of Aarhus, Ny Munkegade 120, DK–8000 Aarhus C, Denmark\\
$^{54}$ Perimeter Institute for Theoretical Physics, Waterloo, Ontario N2L 2Y5, Canada\\
$^{55}$ Department of Physics and Astronomy, University of Waterloo, Waterloo, Ontario N2L 3G1, Canada\\
$^{56}$ Centre for Astrophysics, University of Waterloo, Waterloo, Ontario N2L 3G1, Canada\\
$^{57}$ Space Science Data Center, Italian Space Agency, via del Politecnico snc, 00133 Roma, Italy\\
$^{58}$ Institute of Space Science, Bucharest, Ro-077125, Romania\\
$^{59}$ Departamento de Astrof\'{i}sica, Universidad de La Laguna, E-38206, La Laguna, Tenerife, Spain\\
$^{60}$ Instituto de Astrof\'{i}sica de Canarias. Calle V\'{i}a L\`{a}ctea s/n, 38204, San Crist\'{o}bal de la Laguna, Tenerife, Spain\\
$^{61}$ INFN-Sezione di Roma, Piazzale Aldo Moro, 2 - c/o Dipartimento di Fisica, Edificio G. Marconi, I-00185 Roma, Italy\\
$^{62}$ Dipartimento di Fisica e Astronomia “G.Galilei", Universit\'a di Padova, Via Marzolo 8, I-35131 Padova, Italy\\
$^{63}$ Infrared Processing and Analysis Center, California Institute of Technology, Pasadena, CA 91125, USA\\
$^{64}$ Instituto de Astrof\'isica e Ci\^encias do Espa\c{c}o, Faculdade de Ci\^encias, Universidade de Lisboa, Tapada da Ajuda, PT-1349-018 Lisboa, Portugal\\
$^{65}$ Departamento de F\'isica, Faculdade de Ci\^encias, Universidade de Lisboa, Edif\'icio C8, Campo Grande, PT1749-016 Lisboa, Portugal\\
$^{66}$ Universidad Polit\'ecnica de Cartagena, Departamento de Electr\'onica y Tecnolog\'ia de Computadoras, 30202 Cartagena, Spain\\
$^{67}$ Kapteyn Astronomical Institute, University of Groningen, PO Box 800, 9700 AV Groningen, The Netherlands
}
\date{Received \today \ / Submittes --}
	
\abstract{The \Euclid\ satellite, to be launched by ESA in 2022, will be a major instrument for cosmology for the next decades. \Euclid\ is composed of two instruments: the Visible (VIS) instrument and the Near Infrared Spectromete  and Photometer (NISP). In this work we estimate the implications of correlated readout noise in the NISP detectors for the final in-flight flux measurements. Considering the multiple accumulated (MACC) readout mode, for which the UTR (Up The Ramp) exposure frames are averaged in groups, we derive an analytical expression for the noise covariance matrix between groups in the presence of correlated noise. We also characterize the correlated readout noise properties in the NISP engineering grade detectors using long dark integrations. For this purpose, we assume a $(1/f)^{\, \alpha}$-like noise model and fit the model parameters to  the data, obtaining typical values of  $\sigma = 19.7^{+1.1}_{-0.8}$ e$^{-} \rm{Hz}^{-0.5}$, $f_{\rm{knee}} = \paren{5.2^{+1.8}_{-1.3}} \times 10^{-3} \, \rm{Hz}$ and $\alpha = 1.24 ^{+0.26}_{-0.21}$.  
Furthermore, via realistic simulations and using a maximum likelihood flux estimator we derive the bias between the input flux and the recovered one. We find that using our analytical expression for the covariance matrix of the correlated readout noise we diminish this bias by up to a factor of four with respect to the white noise approximation
for the covariance matrix. Finally, we conclude that the final bias on the in-flight NISP flux measurements should still be negligible even in the white noise approximation, which is taken as a baseline for the \Euclid\ on-board processing.
 }
\titlerunning{Impact of correlated readout noise for the \Euclid\ NISP instrument.}	
\authorrunning{A. Jim\'enez Mu\~noz  et al.}
\maketitle

\section{Introduction}\label{sec:introduction}

\begin{figure*}[h!] 
    \centering 
            {\includegraphics[width = 0.4\textwidth]{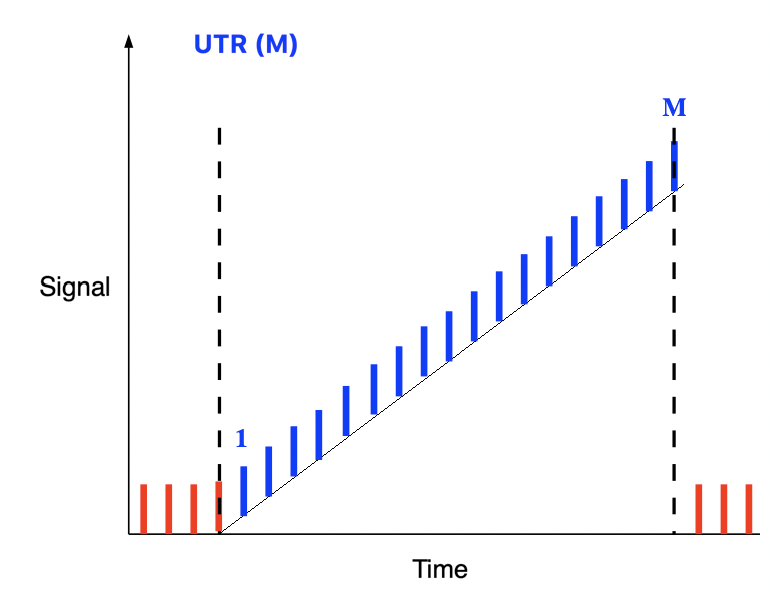}\label{fig:utr}}
            {\includegraphics[width = 0.5\textwidth]{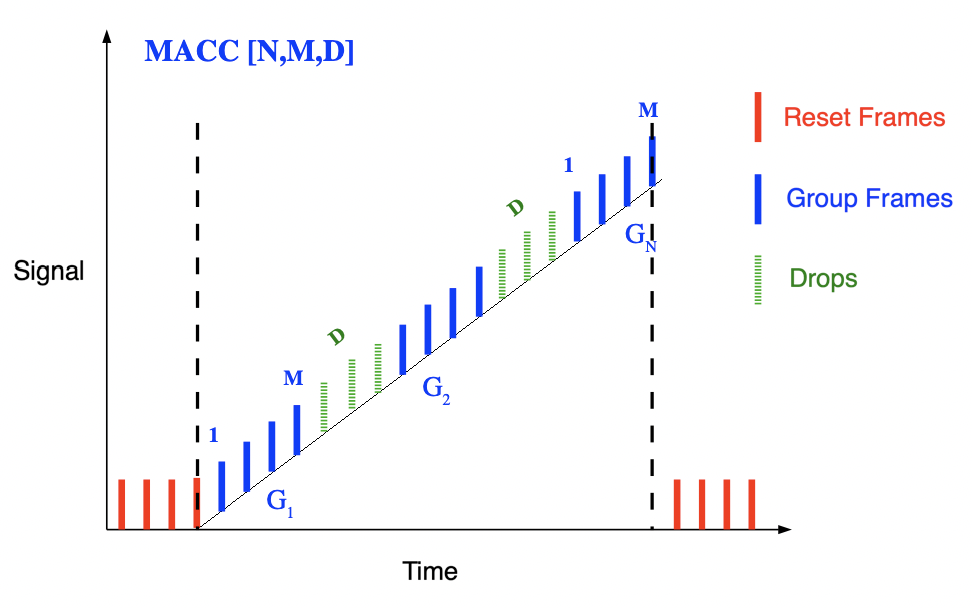}\label{fig:macc}}
            \caption{From left to right, we present a schematic view of the Up the Ramp (UTR), and Multiple Accumulative (MACC) read modes, respectively. The reset, acquired and dropped frames are shown in red, blue and green.}
\label{fig:lecmodes}
\end{figure*}

\indent\indent \Euclid\ \footnote{https://www.euclid-ec.org/} is a Medium Class satellite mission to be launched by ESA in 2022. 
The \Euclid\ satellite is mainly devoted to cosmology and intends to unveil the nature of the Dark Energy
and the Dark Matter \citep{2020A&A...642A.191E}. These two components dominate the  content of the Universe today \citep{ refId0} and are responsible for the accelerated expansion of the Universe and large scale structure formation, respectively. 

\Euclid\ will perform a survey of $15\,000$ $\rm{deg}^2$ of the extragalactic sky with two instruments in the visible and near-infrared (NIR) domains \citep[see][for details]{laureijs2012euclid}. The visible instrument, VIS, operates in the visible regime providing high quality images in a wide-band to carry out precise weak lensing galaxy shear measurements. The NIR instrument, NISP (NIR spectrometer and photometer), provides photometric measurements in three NIR  bands (\textit{Y}, \textit{J}, and \textit{H}) and slitless spectroscopy in a wide NIR band using blue (920-1250 nm)  and red (1250-1850 nm) grisms. The NISP is designed to provide photometric redshifts of about  2 billion galaxies, as well as spectroscopic redshifts for more than 50 million galaxies. In terms of performance the NISP should be able to observe very faint IR distant galaxies obtaining a signal-to-noise ratio of at least 5 down to magnitudes of 24.0 in about 100 seconds per filter. 
The NISP focal plane  \citep{2016SPIE.9904E..0TM} holds 16 NIR sensitive H2RG (HgCdTe astronomical wide area infrared imager or HAWAII) detectors supplied by Teledyne \citep{2008SPIE.7021E..0HB} and selected by NASA. Each one consists of an array of $2048\times2048$ pixels with $2040 \times 2040$ photosensitive pixels, the remaining ones, called reference pixels, being used for tracking biases and temperature variations over long exposures \citep{2010SPIE.7742E..1BM,2012AAS...21932806R}.


Generally, near-infrared detectors, as those of the NISP, perform non-destructive exposures of hundreds of seconds in an up the ramp (UTR) mode (left panel of Fig.~\ref{fig:lecmodes}). For each exposure, non-destructive frames containing the full array data (i.e., all the pixels) are read (without instrumental reset) at about 1 Hz and sent to the acquisition program. 
The set of frames, for a given exposure, forms a ramp and the total flux for the exposure can be computed for each pixel through the slope of the ramp via a simple linear fit \citep{2012AAS...21932806R,2016PASP..128j4504K}. The final uncertainties in the measured flux are given by the uncertainties in each of the frames, which are due both to the readout noise and the photon noise. In order to reduce the noise in the estimation of the flux the multiple accumulated  (MACC, right panel of Fig.~\ref{fig:lecmodes}) sampling mode is applied by averaging the UTR frames into groups
that are then sent to the acquisition system~\citep{secroun2016characterization}.

In the case of the NISP instrument the typical exposure time varies from about 87 s in photometric mode to 547 s in spectrometric mode, and the frame readout frequency is about $0.692 \, \rm{Hz}$. In flight the MACC mode is used but as the amount of data is too large for the \Euclid\ telemetry the individual groups are not transferred to Earth. Indeed, only the flux (ramp fitted slope) for each pixel is available for further data processing. Due to on-board CPU limitations the slopes are obtained from an analytical solution of the linear fit. As a consequence to obtain an accurate flux estimate it is necessary to have an accurate description of the photon and readout noise. In the current \Euclid\ baseline both the readout and photon noise are described by a white noise approximation \citep{fowler1990demonstration,kubik2015optimization}. The main properties of the readout noise assuming white noise are characterized during ground calibration and used in-flight.
However, it has been found that individual H2RGs may present some level of correlated noise in the form of $(1/f)^{\, \alpha}$-like noise~\citep{SMADJA2010288,KUBIK2015315}. Such noise correlation might bias \Euclid\ in-flight flux estimates. 

In this paper we use ground calibration data to characterize the readout noise of the NISP detectors in terms of correlated $(1/f)^{\, \alpha}$-like noise
and study possible bias in the flux measurements induced by departures from the white noise approximation in the linear fit.
The paper is structured as follows. In Sect.~\ref{sec:flux_estimation} analytical expressions for the noise covariance matrix of the MACC groups and for the maximum likelihood solution of the linear fit in the case of correlated readout noise are derived. Section~\ref{sec:fit} describes the characterization of the readout noise for the calibration data of the NISP detectors in terms of correlated noise.  In Sect.~ \ref{sec:discussions_bias}, we estimate the expected flux bias in the case of realistic correlated readout noise. 
Finally, we conclude in Sect.~\ref{sec:conclusion}.

\section{Flux estimation in the case of correlated readout noise}\label{sec:flux_estimation}

\subsection{Flux estimation}\label{sec:covariance}

\indent\indent  Here we concentrate on the MACC readout mode, for which the acquired data consist of $n$ groups with $m$ frames each, and $n-1$ drops (non acquired frames)\footnote{Drops are necessary to reduce the processing time on-board.} with $d$ frames each: $\rm{MACC}(n,m,d)$. During flight EUCLID uses $\rm{MACC}(4,16,4)$ and $\rm{MACC}(15,16,11)$ for photometry and spectroscopy, respectively.
Following \citet{kubik2015optimization} we estimate the total flux from the group differences,
$\Delta G_{k} = G_{k+1} - G_{k}$.

For each group $k$, we define $S_i^{(k)}$ as the total signal in a frame, $f_0$ as the signal accumulated between the last reset and the first readout of the pixel, and $f_i^{(k)}$ as the signal between the frames $i^{(k)}$ and ${(i+1)}^{(k)}$. Furthremore, we notate $\rho_i^{(k)}$ as the readout noise of the frame $i^{(k)}$, and, we represent the non-transferred signal between the last frame of a group and the first frame of the next one, which is accumulated in the drops, by $\mathcal{D}^{(k)}$. Thus, accounting for signal and readout noise contributions, $G_{k}$, the averaged measured signal for group $k$, is given by

\begin{equation}
\begin{aligned}
G_k &= \frac{1}{m}\sum_{i=1}^m S_i^{(k)} = \frac{1}{m}\sum_{i=1}^m \rho_i^{(k)}+\frac{1}{m}\sum_{i=1}^{m-1} (m-i)f_i^{(k)} \\
& + f_0 +\sum_{j=1}^{k-1}\left[\sum_{i=1}^{m-1} f_i^{(j)} + \,\mathcal{D}^{(j)}\right]
\end{aligned}
\end{equation}

and then

\begin{equation}
\begin{aligned}
G_{k+1}-G_k &= 
\,\mathcal{D}^{(k)}+\frac{1}{m}\sum_{i=1}^{m-1}\left[ if_{i}^{(k)}+(m-i)f_i^{(k+1)}\right] \\
&+ \frac{1}{m}\sum_{i=1}^{m}\left(\rho_i^{(k+1)}-\rho_i^{(k)}\right).
\end{aligned}
\label{eq:dif_group_gen}
\end{equation}

\noindent Using these group differences, we can then derive a  Maximum Likelihood estimator for the total flux, $g$. We use the following Gaussian approximation for the likelihood function

\begin{equation}
\begin{aligned}
\mathscr{L} = \dfrac{1}{\sqrt{(2\pi)^{(n-1)} \, |D|}} \exp{\left[-\dfrac{1}{2}(\vec{\Delta G}-\vec{g})^T D^{-1}(\vec{\Delta G}-\vec{g})\right]},
\end{aligned}
\label{eq:likelihood}
\end{equation}

\noindent where $\vec{\Delta G} = \lbrace \Delta G_{k}, \ k=1,n-1 \rbrace$ is a vector gathering all group differences. $D$ is the covariance matrix for the group differences, and $|D|$ is its determinant.

\subsection{Correlated noise covariance matrix}\label{sec:corr_equations}

\indent\indent We discuss here the computation of the group difference noise covariance matrix in the case of correlated readout noise. We refer to \citet{kubik2015optimization} for the white noise case.
We will account for both photon and readout noises.
With respect to photon noise, the flux integrated over a frame is Poisson distributed and stochastically independent between frames.
This applies both to fluxes of frames within a group and within a drop. We can then write:
\begin{equation}
\begin{aligned}
\langle \delta f_i^k \delta f_j^l \rangle =f \delta_{ij}\delta_{kl} \ ;  \\
\langle \delta \mathcal{D}^k \delta \mathcal{D}^l \rangle =\mathcal{D} \,\delta_{kl}.
\end{aligned}
\label{eq:fyd}
\end{equation}

%

\indent The correlated readout noise is assumed to be Gaussian distributed with a $(1/f)^{\, \alpha}$-like spectrum in Fourier domain as in \citet{KUBIK2015315}. In the time domain this is equivalent to a Gaussian distributed noise described by a correlation function\footnote{The correlation function is the inverse Fourier transform of the spectrum.} of the form: $C \left[\left| \delta_t \right| \right]$, where $\delta_t$ is the time interval between two given frames. Thus, we can write
\begin{equation}
\begin{aligned}
\langle\delta\rho_i^k \delta\rho_j^l \rangle = C\left(|(l-k)(m+d)+(j-i)| \, t_{\rm{frame}}\right),
\end{aligned}
\label{eq:correlationfunc}
\end{equation}
\noindent where $t_{\rm{frame}}$ is the frame integration time.

\indent Using Eqs.~(\ref{eq:fyd}) and (\ref{eq:correlationfunc}) the group noise covariance matrix (see Appendix \ref{sec:appendix} for details) reads
\begin{equation}
\begin{aligned}
C_{kk}&=(k-1)\,\mathcal{D}+(k-1)(m-1)f+f\frac{(m+1)(2m+1)}{6m}\\ 
&+\frac{1}{m^2} \left[ mC(0)+2\sum_{i=1}^{m-1}(m-i)C(i \, t_{\rm{frame}}) \right]
\end{aligned}
\label{eq:ckk_corr}
\end{equation}

\noindent for the diagonal terms and 
\begin{equation}
\begin{aligned}
C_{kl}&=(k-1)\,\mathcal{D}+(k-1)(m-1)f+f\frac{(m+1)}{2}\\
&+\frac{1}{m}  \,C \left( (l-k)(m+d) \ t_{\rm{frame}} \right) \\
&+\frac{1}{m^2}
\sum_{i=1}^{m}\sum_{j=1,j\neq i}^{m}C\left( |(l-k)(m+d)+ (j-i)| \, t_{\rm{frame}} \right)
\end{aligned}
\label{eq:ckl_corr}
\end{equation}

\noindent for the off-diagonal ones. By constrast to the white noise case described by \cite{kubik2015optimization} we find in the latter expression a contribution from the readout noise.

We also derive the group difference covariance matrix. The diagonal terms are given by

\begin{equation}
\begin{aligned}
D_{kk}&=\,\mathcal{D}+\frac{(m-1)(2m-1)}{3m}f \\
&+\frac{2}{m^2}\sum_{i=1}^{m}\sum_{j=1}^{m}C\left(|j-i| \, t_{\rm{frame}} \right)\\
&-\frac{1}{m^2}\sum_{i=1}^{m}\sum_{j=1}^{m}C\left(|j-i-m-d|\, t_{\rm{frame}} \right)\\
&-\frac{1}{m^2}\sum_{i=1}^{m}\sum_{j=1}^{m}C\left( |j-i+m+d| \, t_{\rm{frame}} \right)
\end{aligned}
\label{eq:dkk_corr}
\end{equation}
and the off-diagonal terms are
\begin{equation}
\begin{aligned}
D_{kl}&=\left(\frac{m^2-1}{6m}f\right) \, \delta_{l(k+1)} \\
&+\frac{2}{m^2}\sum_{i=1}^{m}\sum_{j=1}^{m}C\left(|(l-k)(m+d)+(j-i)| \, t_{\rm{frame}} \right)\\
&-\frac{1}{m^2}\sum_{i=1}^{m}\sum_{j=1}^{m}C\left(|(l-k-1)(m+d)+(j-i)| \, t_{\rm{frame}} \right)\\
& -\frac{1}{m^2}\sum_{i=1}^{m}\sum_{j=1}^{m}C\left(|(l-k+1)(m+d)+(j-i)| \, t_{\rm{frame}} \right)
\end{aligned}
\label{eq:dkl_corr}
\end{equation}

\noindent for $k<l$. By contrast to the white noise readout noise case presented in equations (7) and (8) of \cite{kubik2015optimization}, we observe that the contribution of the readout noise to the group difference covariance is not constant in the diagonal terms and it adds extra correlation in the off-diagonal ones.

\section{Characterization of the readout noise of the NISP detectors}
\label{sec:fit}

\begin{figure*}[h!] 
    \centering 
      \includegraphics[width=0.49\linewidth,keepaspectratio]{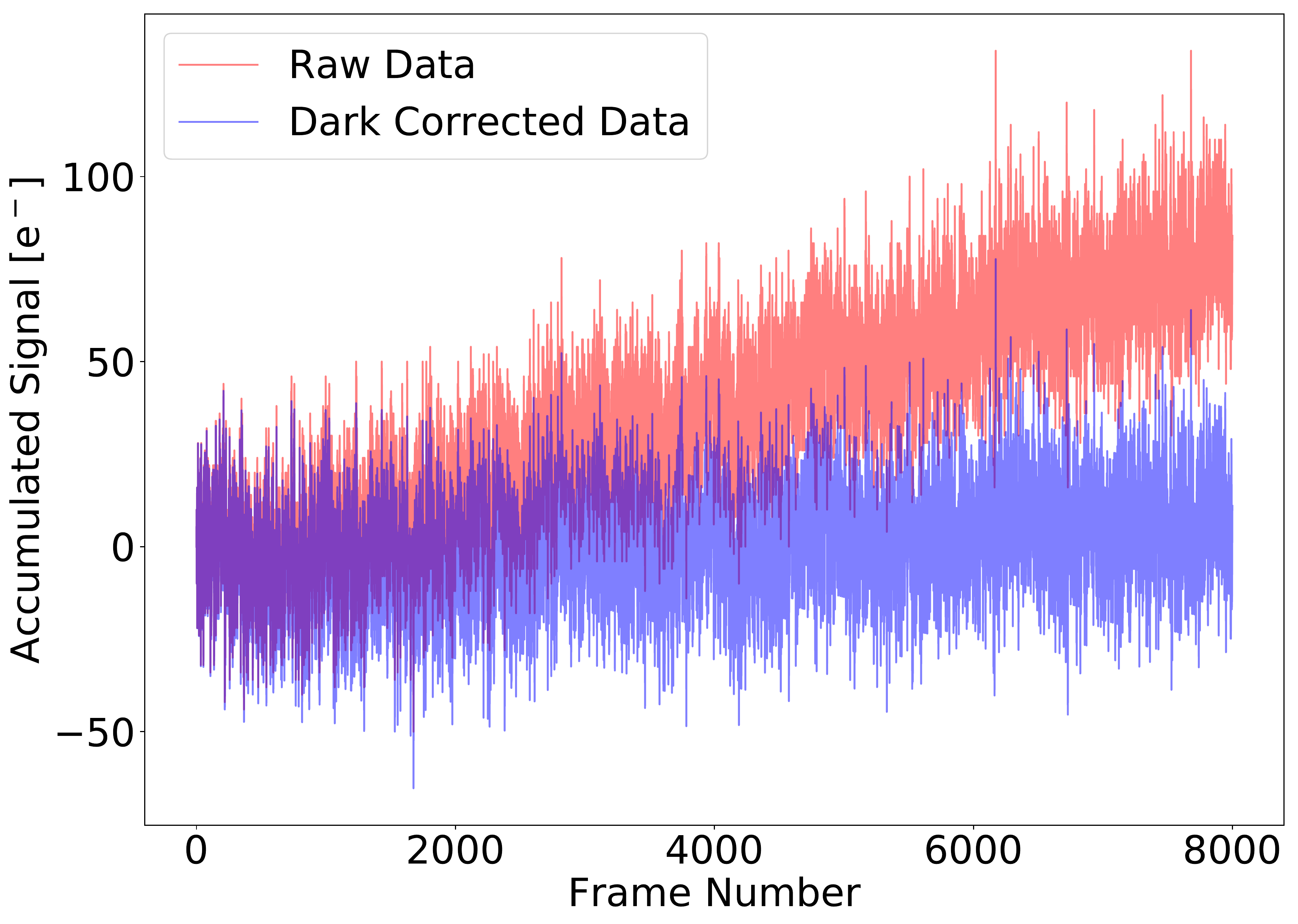} 
 	  \includegraphics[width=0.49\linewidth,keepaspectratio]{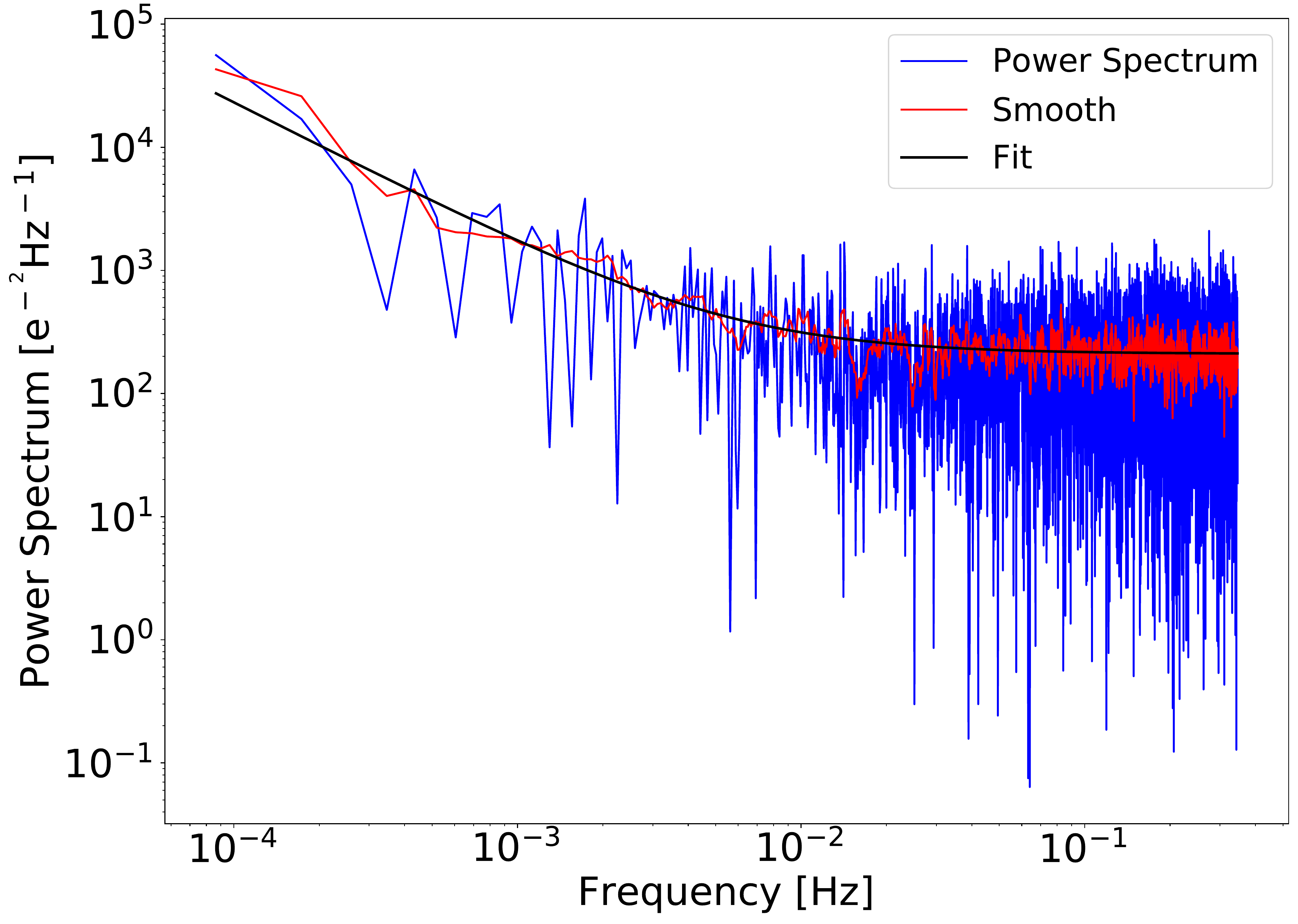}
        \caption{Left panel: Readout noise obtained from the ramp data after dark and pixel reference correction for an inner pixel of the NISP array. Right panel: we show the time power spectrum for the left panel readout noise (blue), the smoothed power spectrum (red) and the best-fit $(1/f)^{\, \alpha}$-like model (black).}
\label{fig:ramp}
\end{figure*}

\subsection{Readout noise measurements}
\indent\indent The readout noise of infrared detectors can be characterized from long exposure ramps in dark conditions. Here, we use dark test data obtained during the \Euclid\ NISP detector characterization performed at the CPPM laboratory. We focus on one of the NISP engineering grade H2RG detectors, which was cooled down to a nominal operating temperature of $85\,$K.  The testing facility was designed to achieve best possible dark conditions and special care was taken to achieve expected in-flight readout noise. 

\indent\indent For proper dark measurement, long integration UTR ramps were acquired, typically, ramps of 8000 frames with a total exposure time of 3.21 hours corresponding to a frame exposure time of $t_{\rm{frame}} = 1.445 \, \rm{s}$. For each frame and for each of the $2040 \times 2040$ photosensitive pixels we use the reference pixels to remove correlations in the readout noise induced by background variations \citep{2010SPIE.7742E..1BM,2012AAS...21932806R}. After reference pixel corrections we find that correlations between pixels are reduced as expected. Furthermore, we compute the dark for each ramp using the \citet{fowler1990demonstration} algorithm, for which the slope of the ramp (in this case the dark contribution) is computed from the difference of the average of blocks of frames at the end and at the beginning of the ramp. In our case we have considered blocks of 32 frames to reduce the uncertainties in the dark measurements. Every ramp is corrected for the dark by subtracting the median dark value of all of the pixels in a given detector.
For the data used in this paper the median dark for all pixels in the array was about $ 6 \times 10^{-3}$ $\rm{e}^{-}\rm{s}^{-1}$ with a standard deviation of $ 2 \times 10^{-3}$ $\rm{e}^{-}\rm{s}^{-1}$ across pixels.
 
The left panel of Fig.~\ref{fig:ramp} shows one of these ramps for one of the inner pixels in the array after correcting for the reference pixels (raw data, red line) and after subtraction of the dark contribution (dark corrected, violet line). We have used a conversion gain factor of $\rm{f}_{\rm{e}} = 2$ $\rm{e}^{-} \rm{ADU}^{-1}$. We can observe in the figure that the readout noise is not fully white. This can be better seen in the right panel of Fig.~\ref{fig:ramp}, where we show the power spectrum of the dark corrected data as a function of the time frequency in Hz.

\subsection{Readout noise modelling and fitting}
The readout noise power spectrum shows a $(1/f)^{\, \alpha}$-like spectrum with an excess of power at low frequencies. Similar patterns are found for all other pixels in the array.
Thus, to characterize the correlated readout noise in the NISP detectors we assume that its power spectrum is given by 
\begin{equation}
\begin{aligned}
P(f) = \frac{\sigma^2}{2} \  \left[ 1 + \left(f_{\rm{knee}}/f \right)^{\, \alpha} \right]
\end{aligned}
\end{equation}
with $f$ the time frequency and $\sigma$, $f_{\rm{knee}}$ and $\alpha$ the parameters of the model. The $\sigma$ parameter gives us information about the flat part of the power spectrum, which corresponds to the white noise contribution. $\alpha$ and $f_{\rm{knee}}$ inform us about the correlated noise contribution.
Notice that for $\alpha =0$ this model converges to a flat power spectrum and thus to a white noise spectrum.

For each pixel in the array we fit the readout noise power spectrum to this $(1/f)^{\, \alpha}$-like model. We use the python {\it mpfit} module, which gives the best-fit parameters and their uncertainties. Uncertainties on the data power spectrum are computed assuming Gaussian noise: $ \sigma_{P(f)} \propto P(f)$. In practice, the fit is performed in two steps. First, 
we estimate the uncertainties in the power spectrum from a smoothed version of the readout noise power spectrum (see red line in the right panel of Fig.~\ref{fig:ramp}) and compute the best-fit parameters for the $(1/f)^{\, \alpha}$-like model. Then, we use these first estimates of best-fit parameters to estimate the uncertainties in the power spectrum and perform a second fit to the readout noise power spectrum. The best-fit parameters obtained from this second fit are stored for further analysis. Using Monte Carlo simulations we have observed that this two-step procedure leads to non biased estimates of the best-fit parameters for the$(1/f)^{\, \alpha}$-like model. In Fig.~\ref{fig:ramp} we show the best-fit $(1/f)^{\, \alpha}$-like model (black line) to the readout noise power spectrum (blue line) obtained from the second fit. 
The best fit-parameters and their uncertainties for this pixel are $\sigma = 20.50\pm 0.23$ $\rm{e}^{-}\rm{Hz}^{-0.5}$, $f_{\rm{knee}} = \paren{5.5 \pm 0.8} \times 10^{-3} \, \rm{Hz}$ and $\alpha = 1.17 \pm 0.15$.
We observe 
that the best-fit model is consistent with the data with $
\chi^2/\rm{N}_{\rm{dof}}$ of 1.57 and $\rm{N}_{\rm{dof}}=8000-3$.

\subsection{Readout noise properties}

\begin{figure*}[h!] 
  \centering
    \subfloat[]{\includegraphics[width=0.38\linewidth,keepaspectratio]{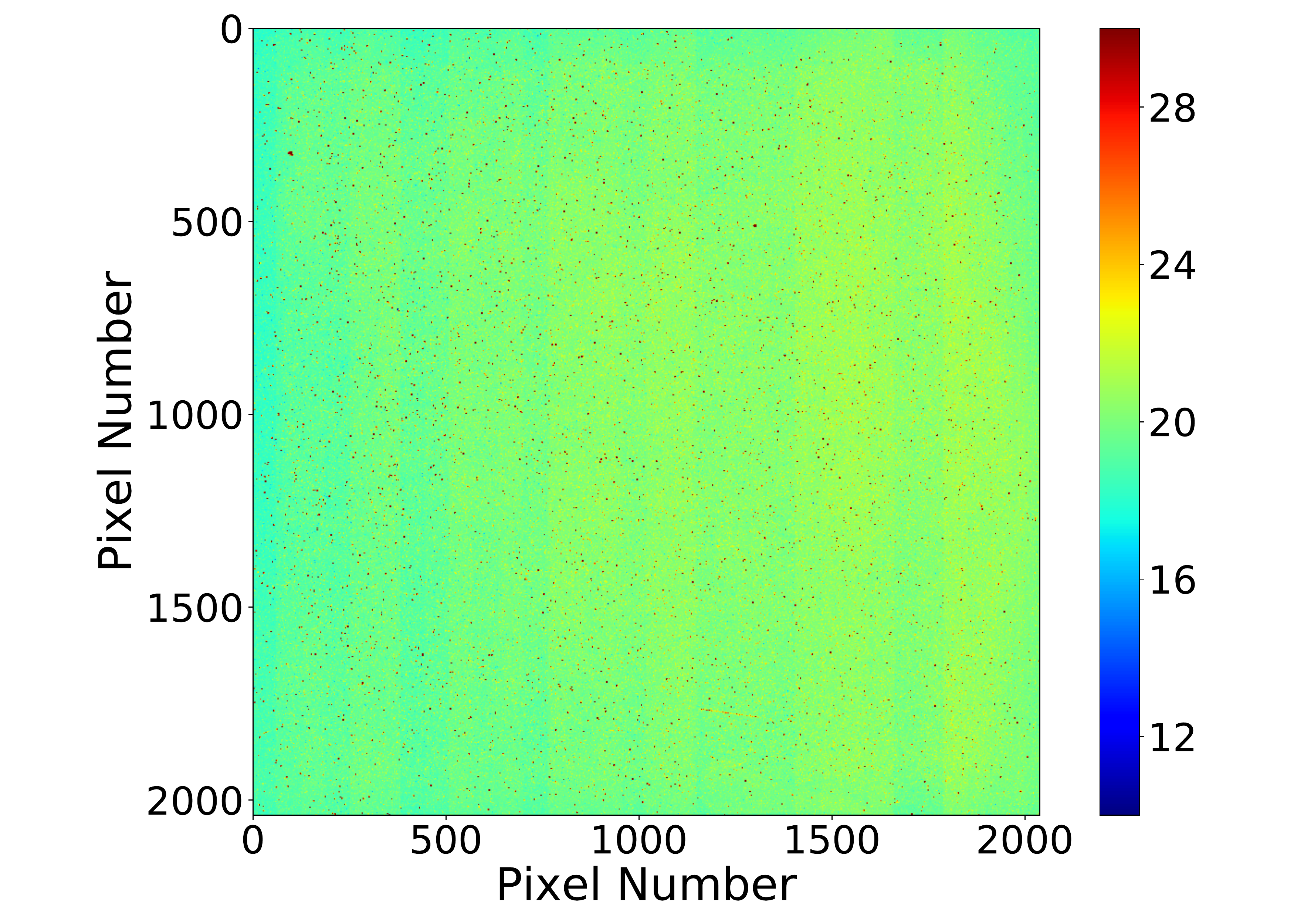}\label{fig:map_sigma}}
             \subfloat[]{ \includegraphics[width=0.38\linewidth, height=0.20\textheight]{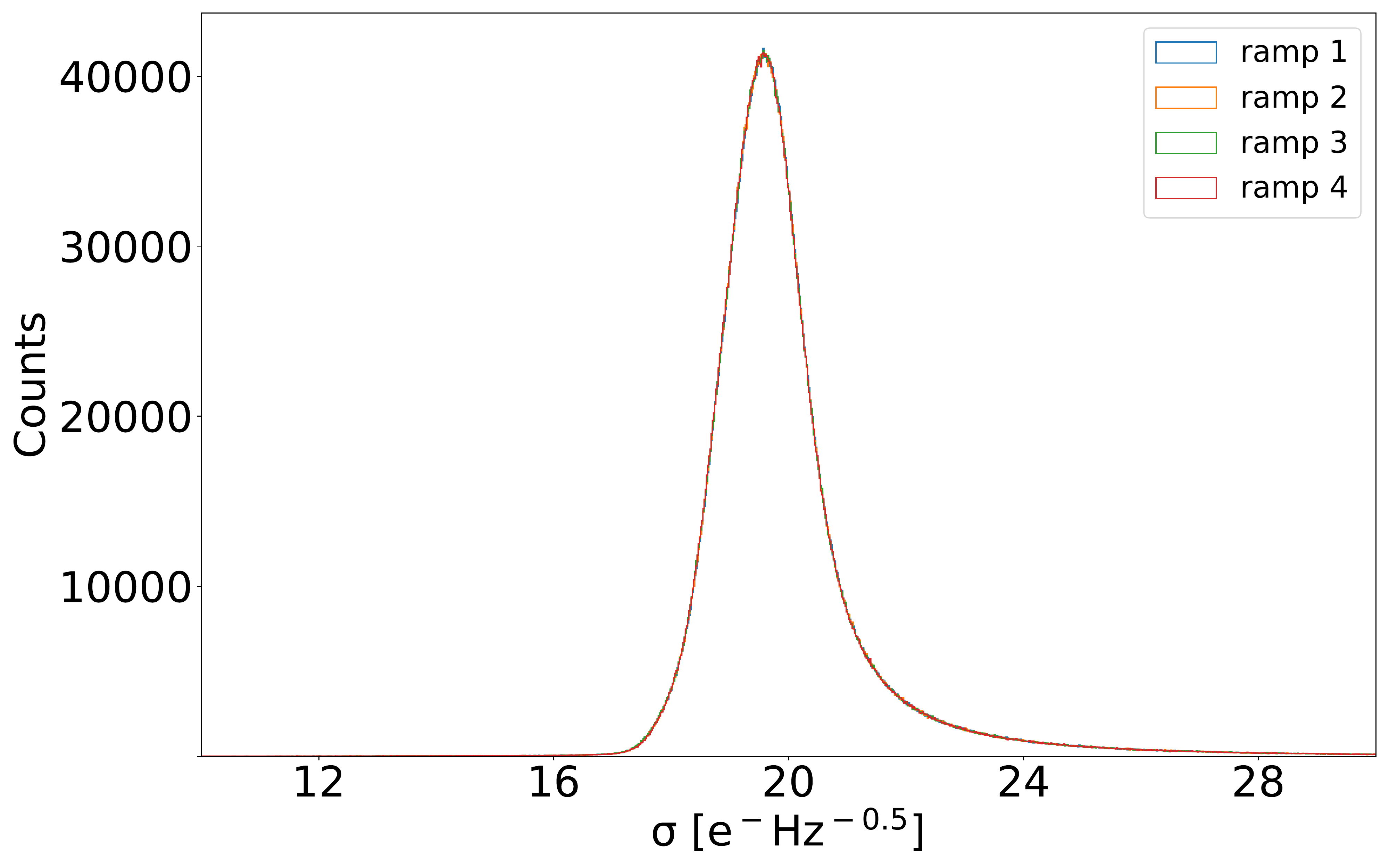}}
               
             \subfloat[] { \includegraphics[width=0.38\linewidth,keepaspectratio]{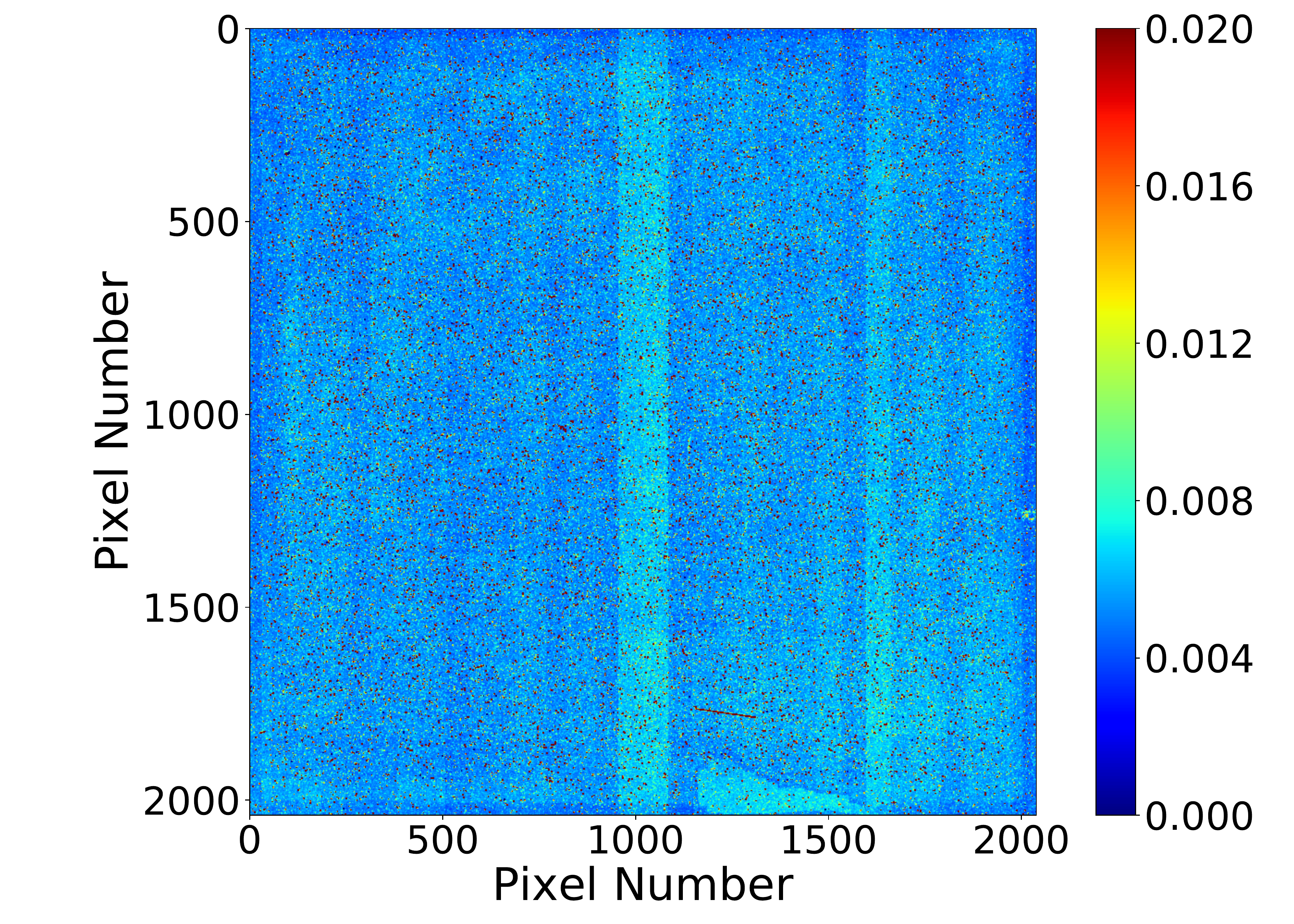}}
 	    		 \subfloat[]{ \includegraphics[width=0.38\linewidth, height=0.20\textheight]{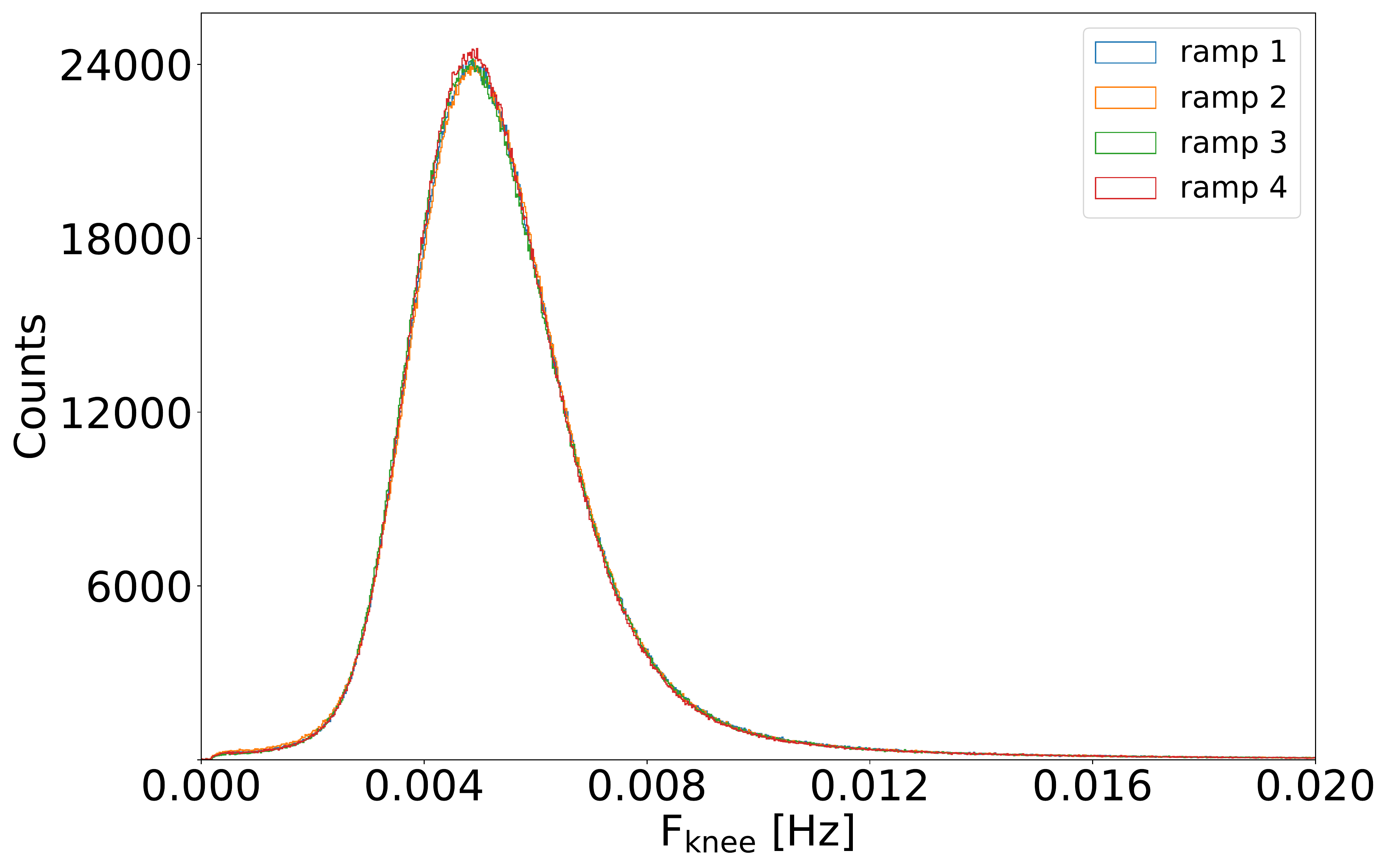}}
 	    		             
         	 \subfloat[]{ \includegraphics[width=0.38\linewidth,keepaspectratio]{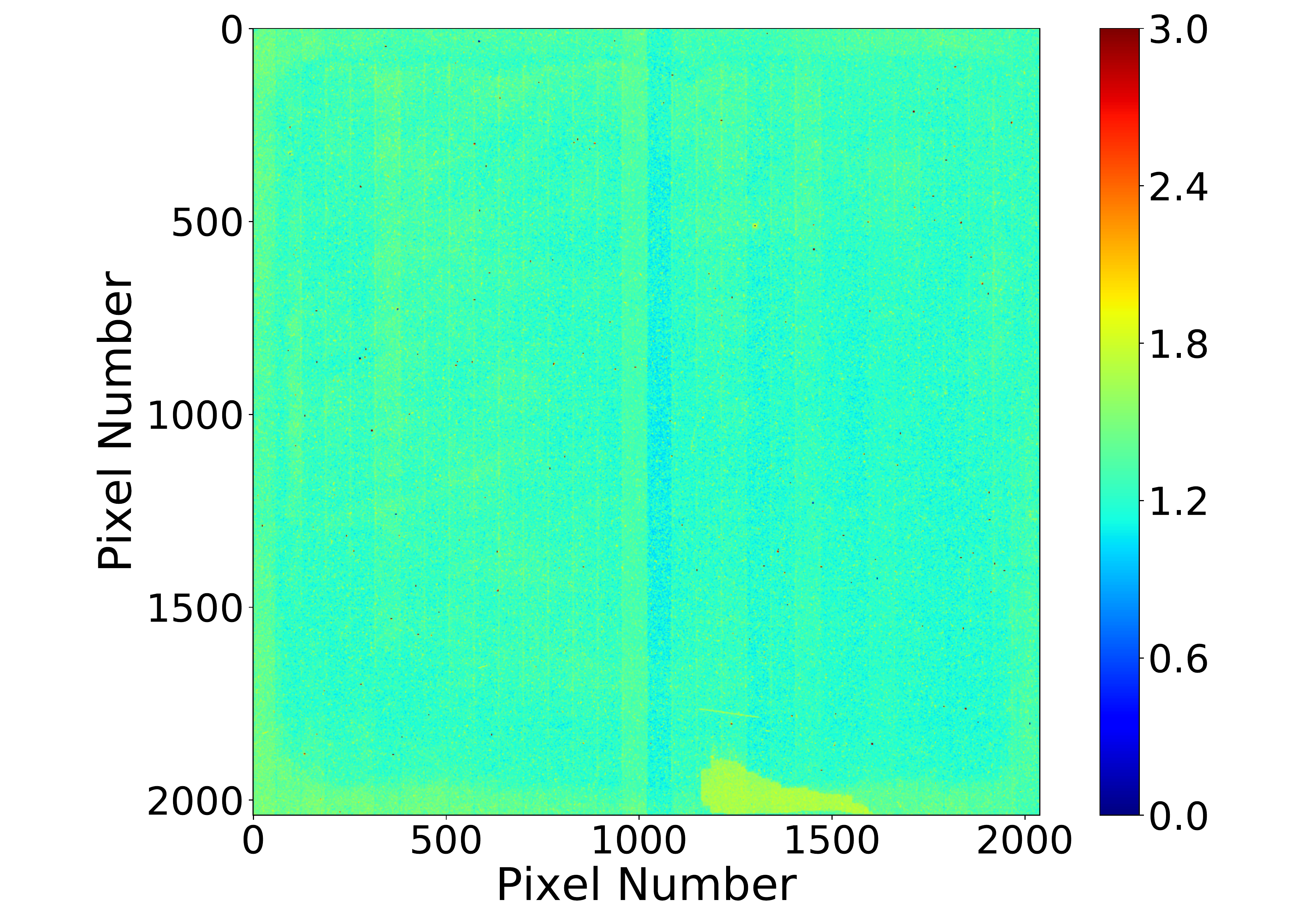}}
      	     \subfloat[]{ \includegraphics[width=0.38\linewidth,, height=0.20\textheight]{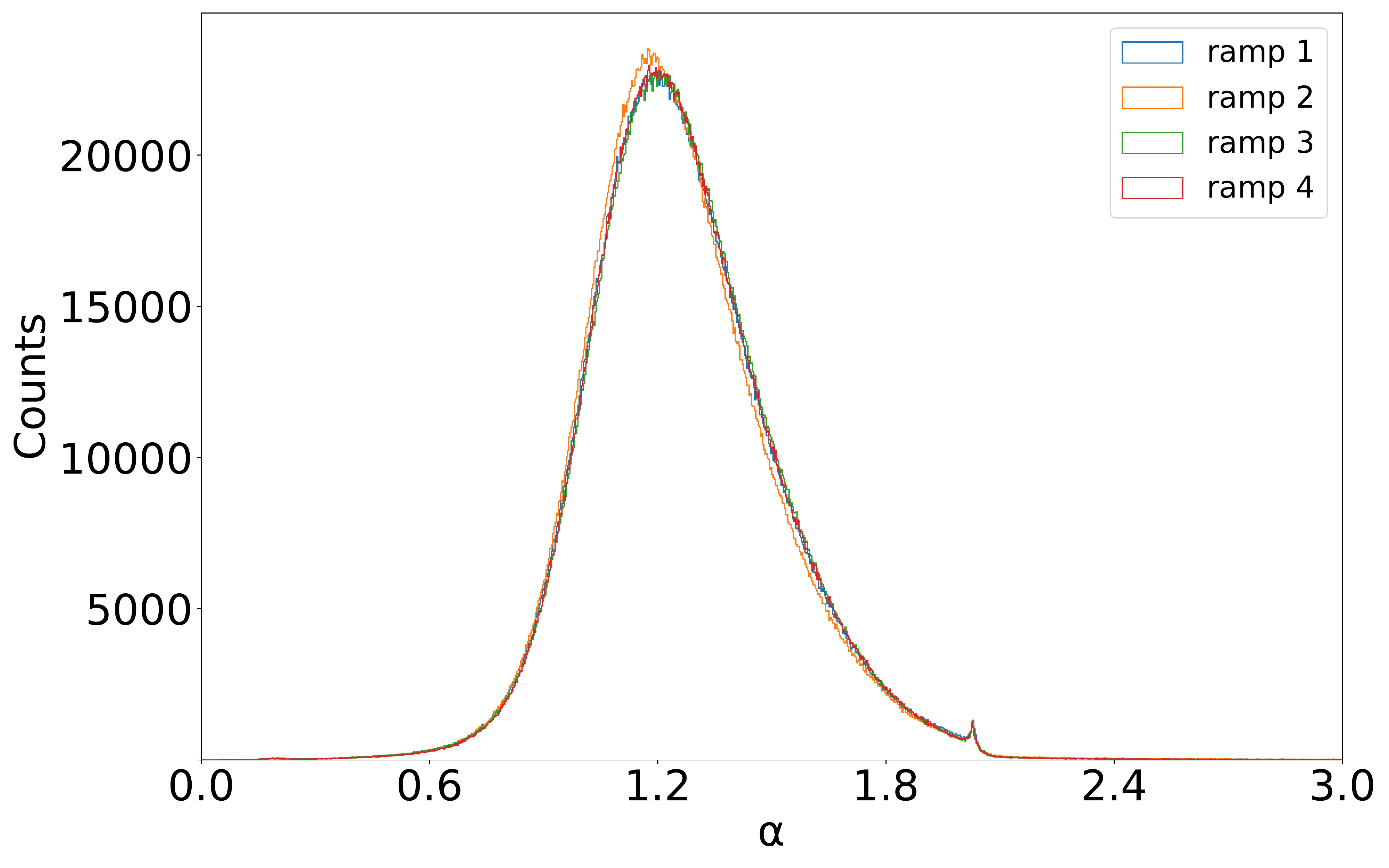}} 
      	
         	 \subfloat[]{ \includegraphics[width=0.38\linewidth,keepaspectratio]{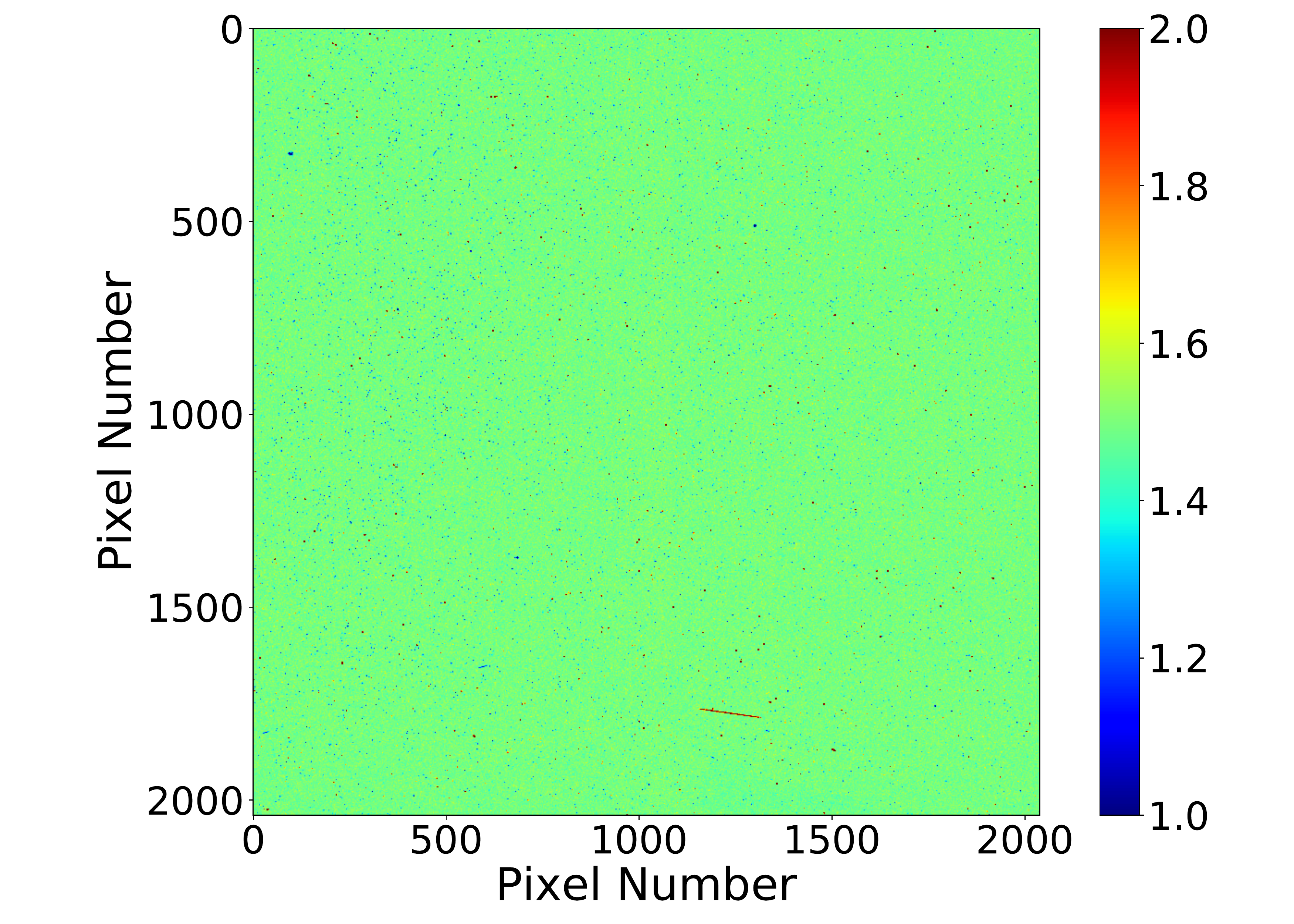}} 
      	     \subfloat[]{ \includegraphics[width=0.38\linewidth,, height=0.20\textheight]{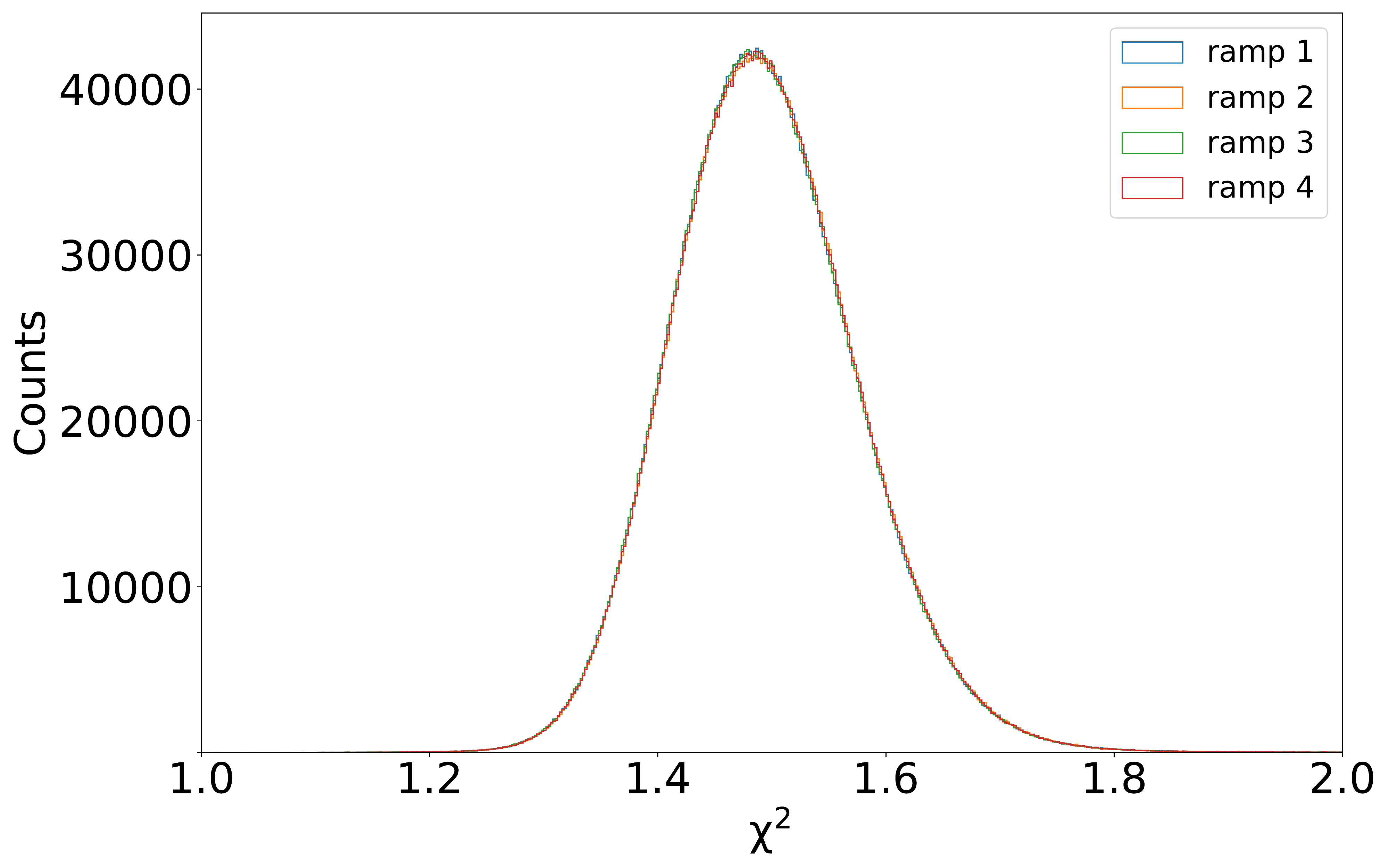}}    

\caption{$(1/f)^{\, \alpha}$-like model best-fit parameters $\sigma$, $f_{\rm{knee}}$ and $\alpha$ for the  photosensitive pixels of one the NISP detectors tested. Panels (a), (c), (e) and (g) are maps representing the best-fit values for the three parameters and the $\chi^2/\rm{N}_{\rm{dof}}$ for the $2040\times2040$ photosensitive pixels of the detector for a single ramp. Panels (b), (d), (f) and (h) show the 1D distribution for the same best-fit parameters of 4 ramps of the same detector, and the $\chi^2/\rm{N}_{\rm{dof}}$, respectively. See main text for details.}
\label{fig:param_maps}
\end{figure*}

We present in the left column of Fig.~\ref{fig:param_maps} four maps representing the best-fit parameters and $\chi^2/\rm{N}_{\rm{dof}}$ values for all the $2040 \times 2040$ photosensitive pixels for one of the ramps of one of the tested detectors. The white dots in the maps correspond to either hot pixels or pixels for which we obtain a bad fit to the data. These pixels represent less than $0.1\%$ of the total pixels and are uniformly distributed in the maps. We can observe in the maps vertical bands which are related to the 32 readout channels, for which we expect some correlations in the noise properties. We can also isolate some particular regions as the one in the $f_{\rm{knee}}$ map for pixels around (2000,1400), which are also found when computing other characteristic quantities of the detectors as for instance the Correlated Double Sampling (CDS) noise. They mainly correspond to manufacturing defects. 

The 1D distributions of the best-fit parameters and the $\chi^2/\rm{N}_{\rm{dof}}$ are shown in the right panels of Fig.~\ref{fig:param_maps} excluding hot and bad-fit pixels. We show in the figure four ramps of the same detector, for which we find consistent results. We observe that the distributions for the three parameters are skewed towards large values. We find that the median values for the best-fit parameters are $\sigma = 19.7^{+1.1}_{-0.8}$ $\rm{e}^{-} \rm{Hz}^{-0.5}$, $f_{\rm{knee}} = \paren{5.2^{+1.8}_{-1.3}} \times 10^{-3} $ $\rm{Hz}$ and $\alpha = 1.24 ^{+0.26}_{-0.21}$. 

\begin{figure*}[h!] 
    \centering 

             \subfloat[]{\includegraphics[width = 0.47\textwidth]{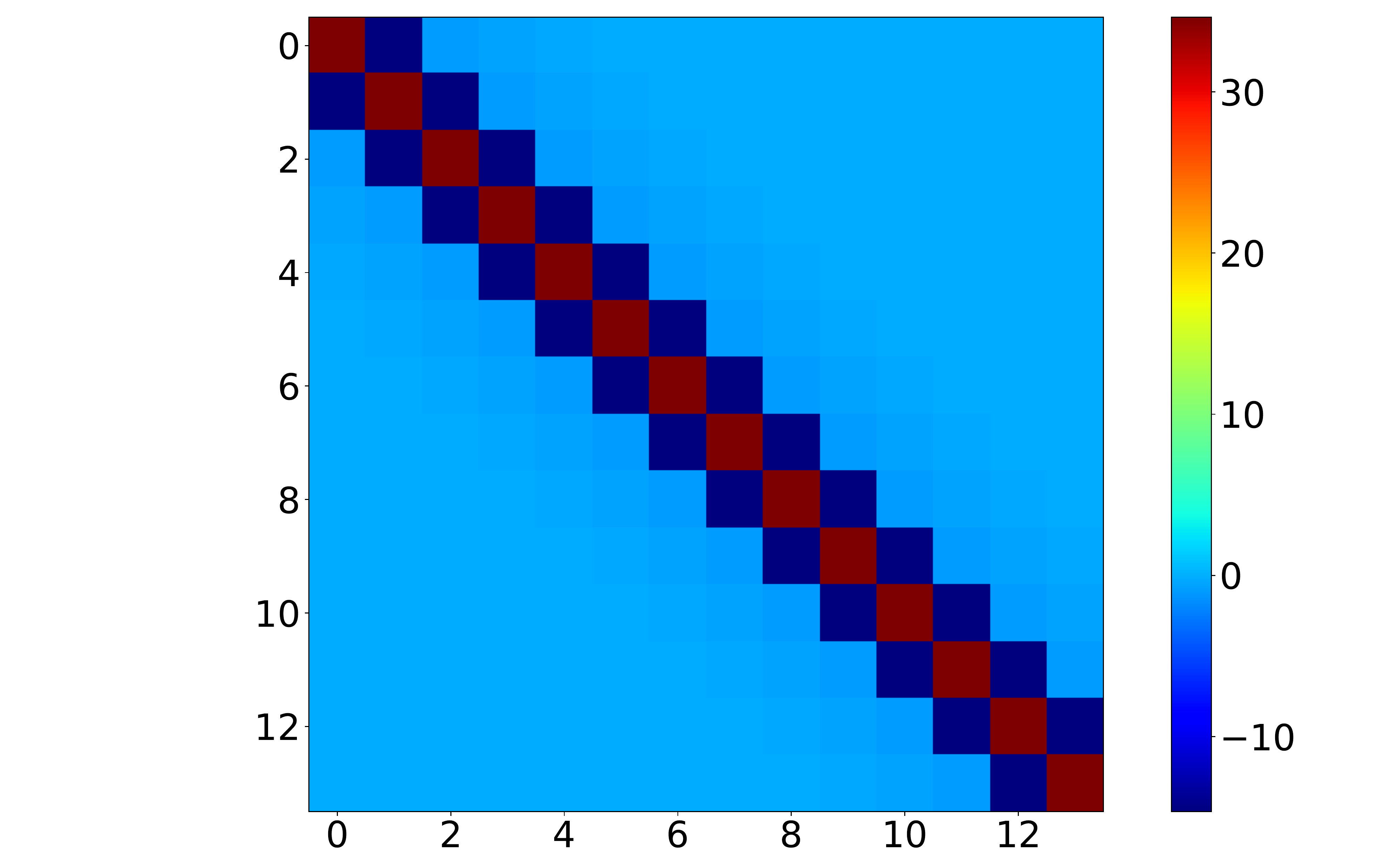}\label{fig:cov_corr_theo}}\hfill  
            \subfloat[]{\includegraphics[width = 0.47\textwidth]{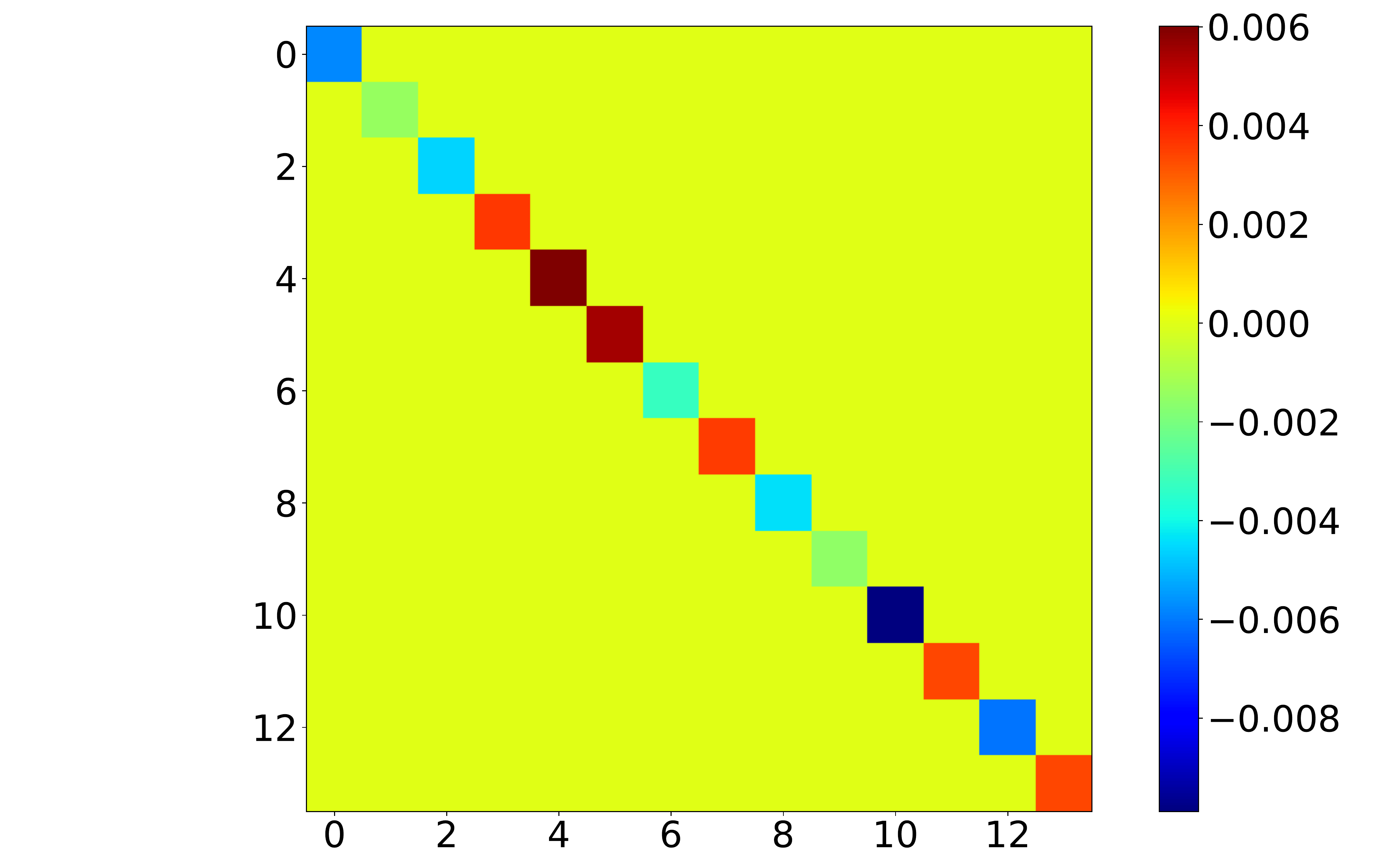}\label{fig:cov_corr_mcmc}}
\caption{Group difference covariance matrix as obtained analytically from Eqs.~(\ref{eq:dkk_corr}) and (\ref{eq:dkl_corr}) (panel a) and relative difference with respect to the one obtained from Monte Carlo simulations (panel b). See text for details.}
\label{fig:cov_matr_comparison}
\end{figure*}

\begin{figure*}[h!] 
    \centering 
            \subfloat[]{\includegraphics[width = 0.32\textwidth]{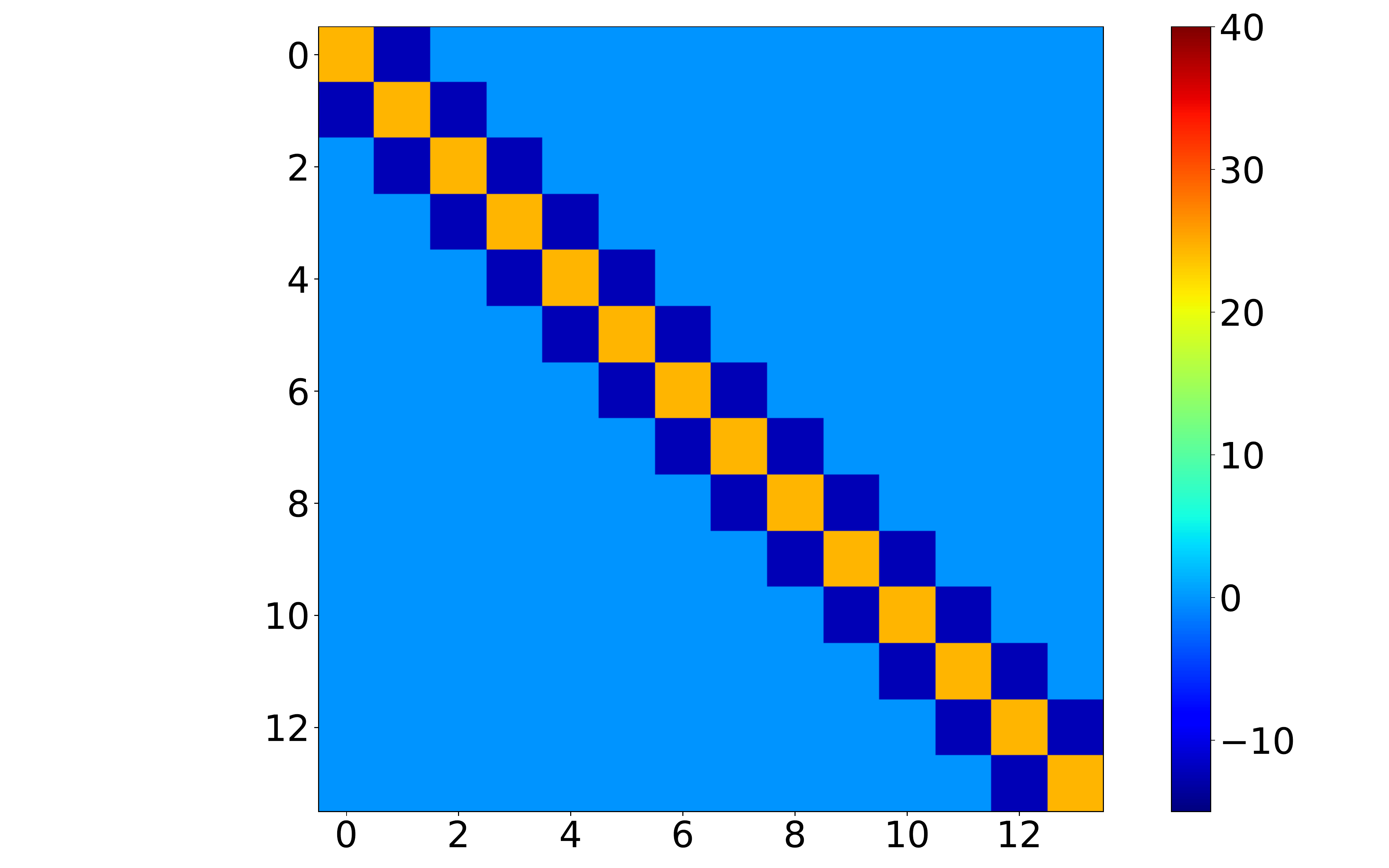}\label{fig:cov_white_lowdark}}            
             \subfloat[]{\includegraphics[width = 0.32\textwidth]{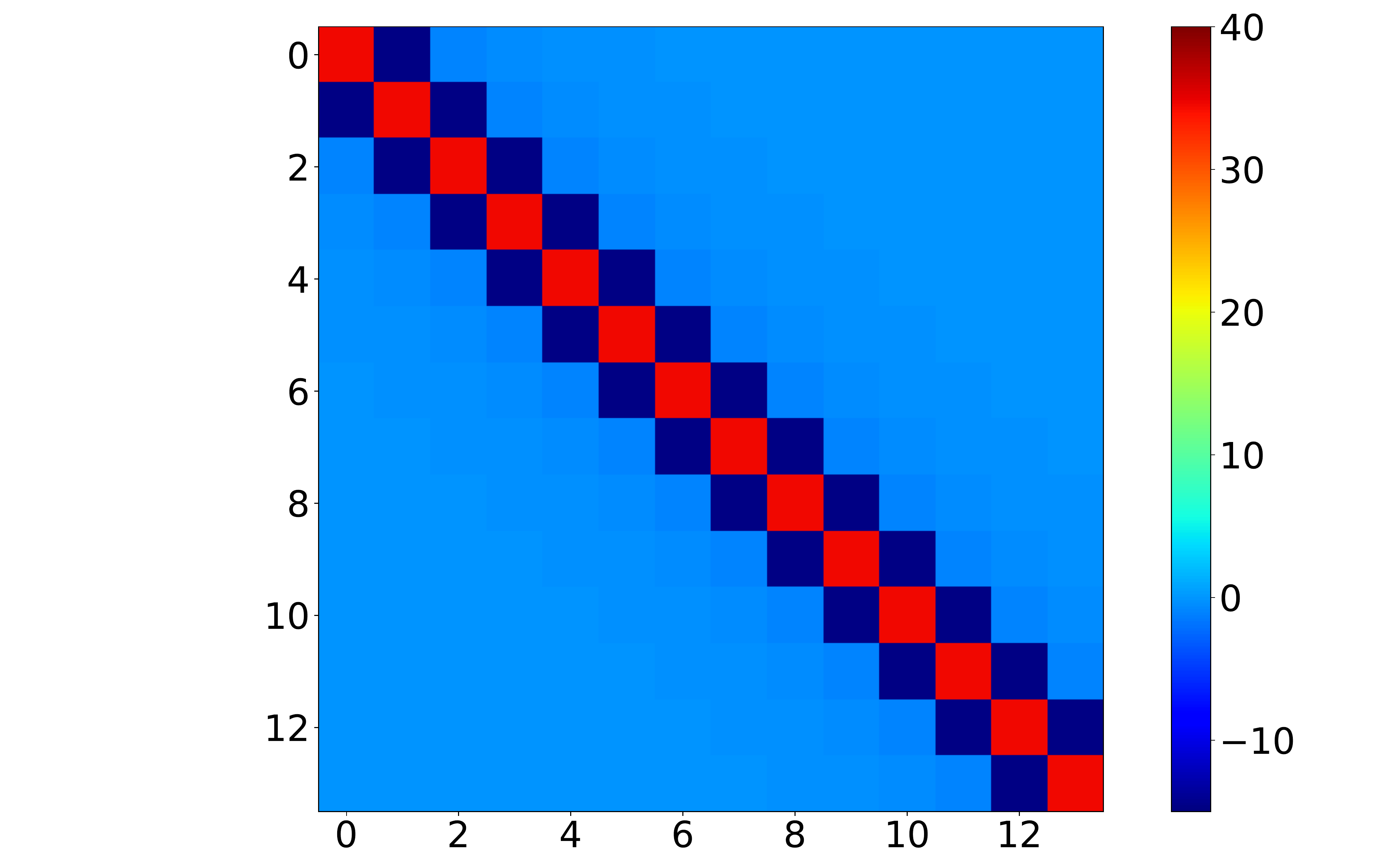}\label{fig:cov_mcmc_lowdark}}
             \subfloat[]{\includegraphics[width = 0.32\textwidth]{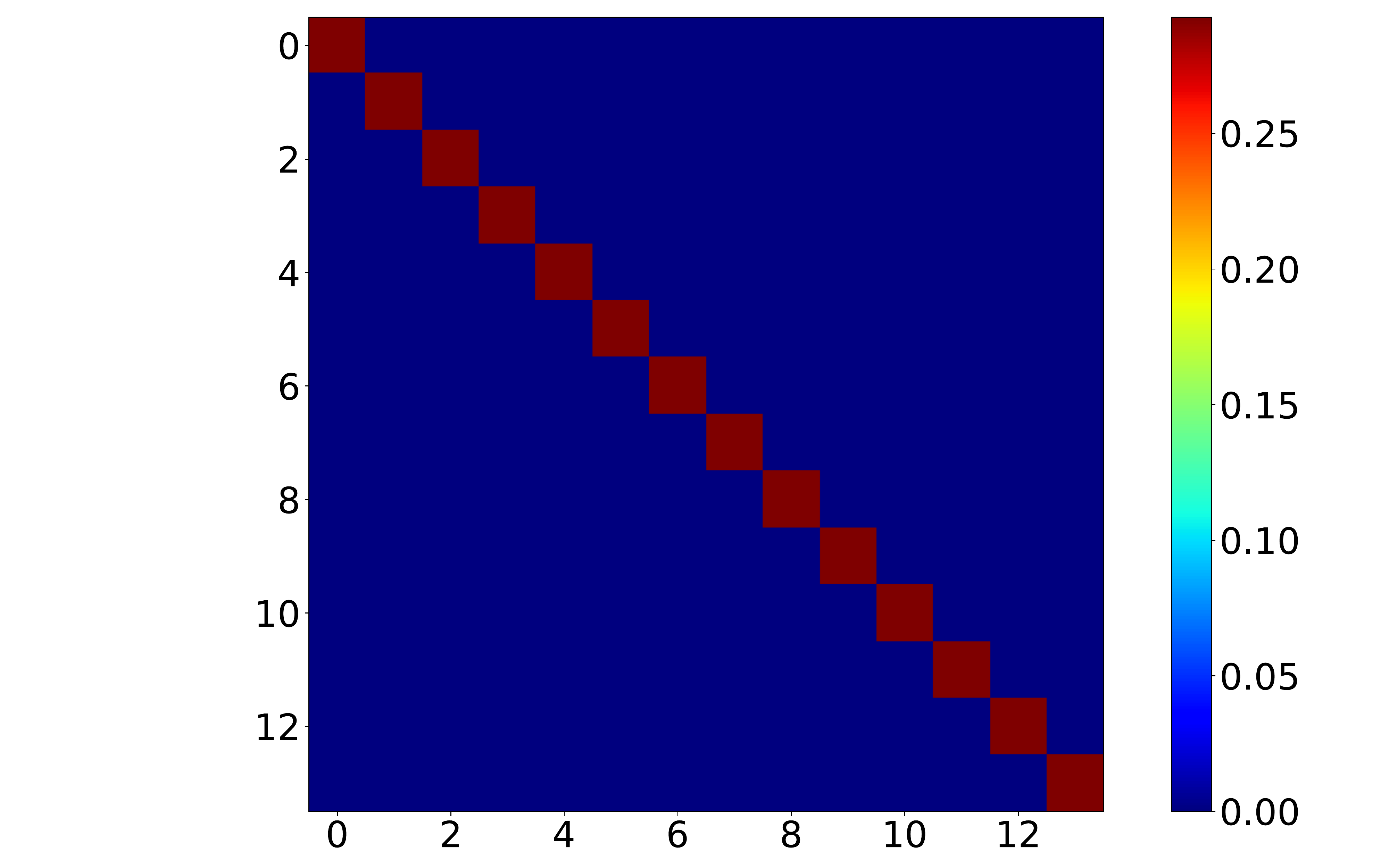}\label{fig:cov_reldiff_lowdark}}
         	              
            \subfloat[]{\includegraphics[width = 0.32\textwidth]{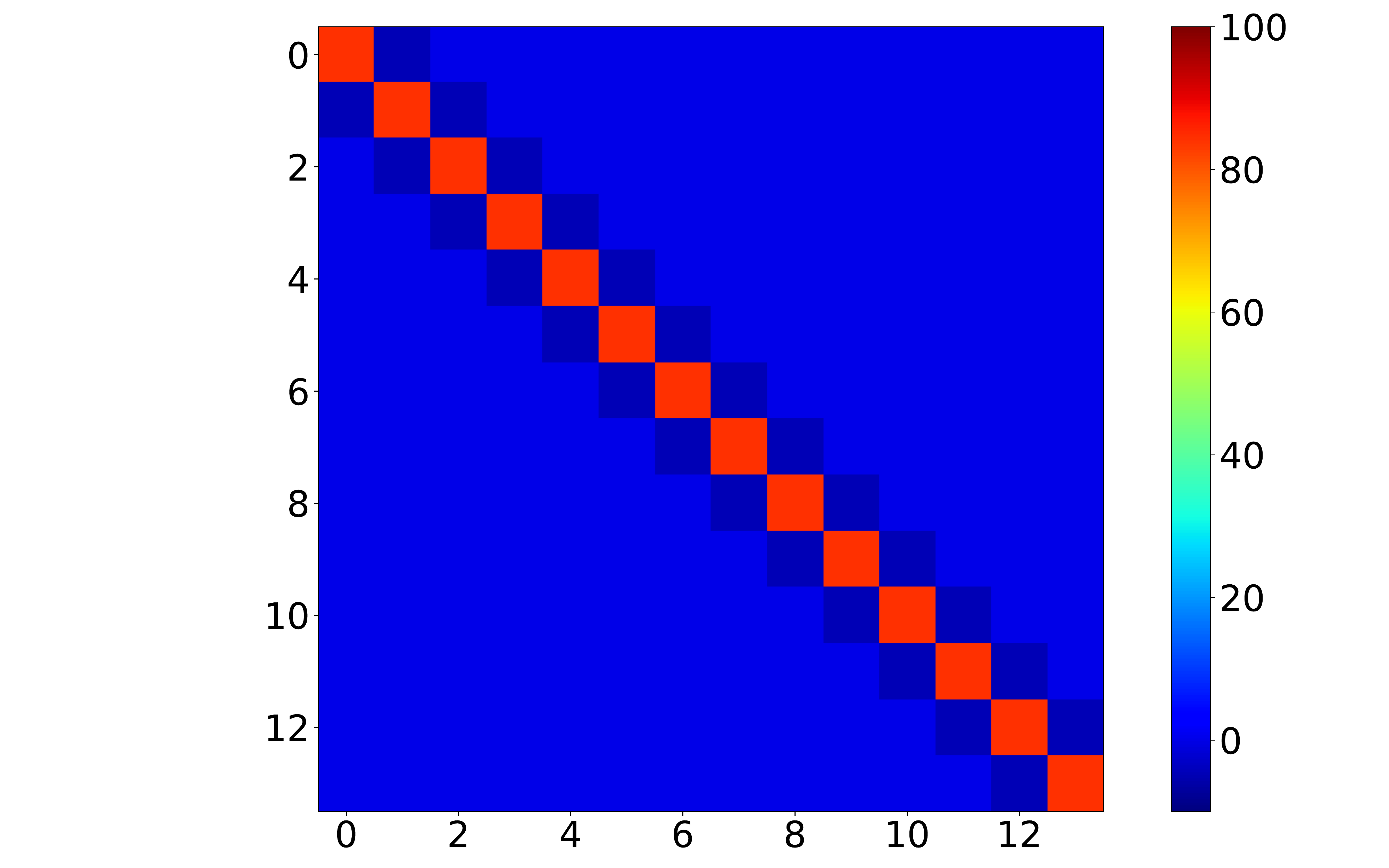}\label{fig:cov_white_highdark}}  
             \subfloat[]{\includegraphics[width = 0.32\textwidth]{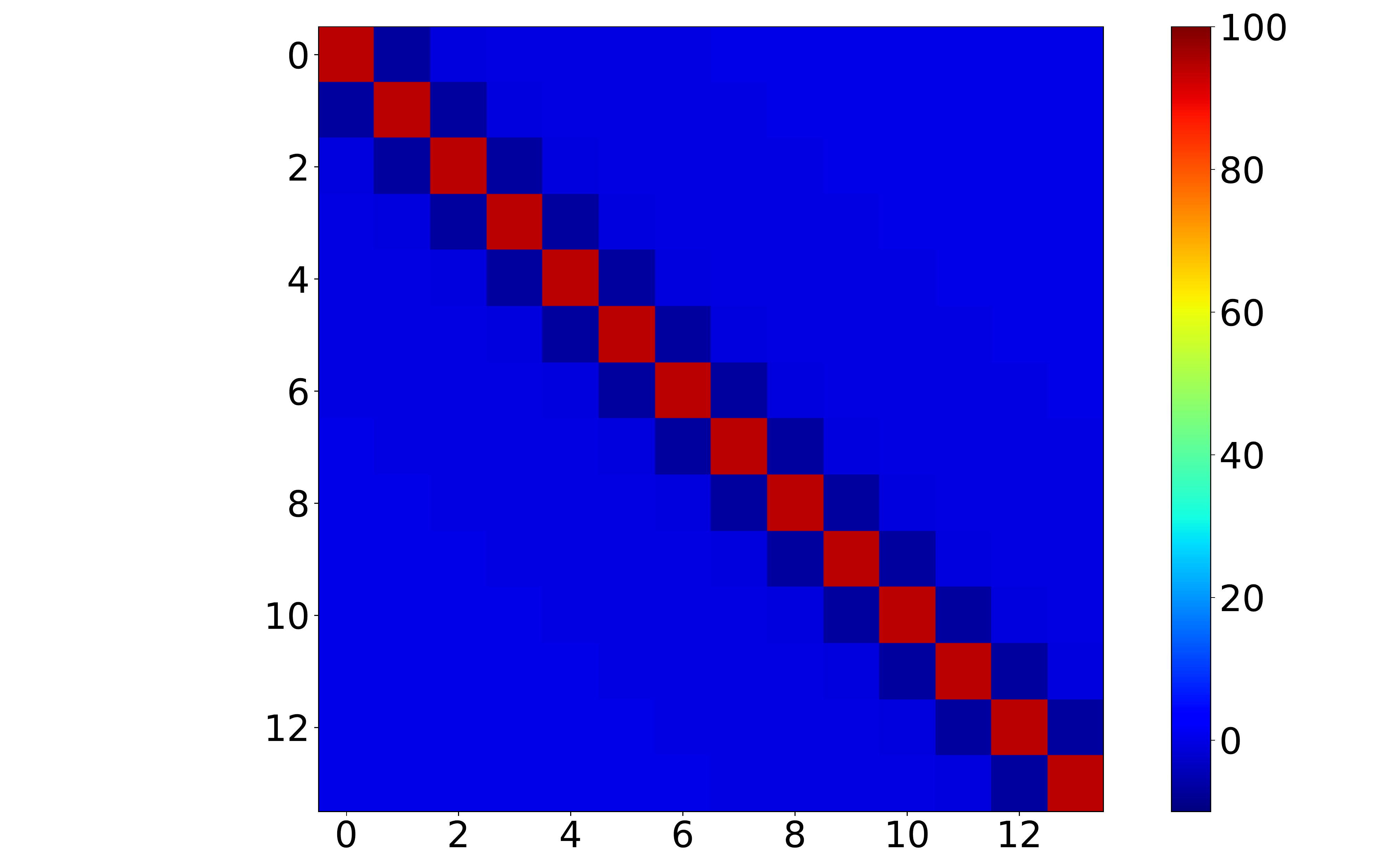}\label{fig:cov_mcmc_highdark}}           
             \subfloat[]{\includegraphics[width = 0.32\textwidth]{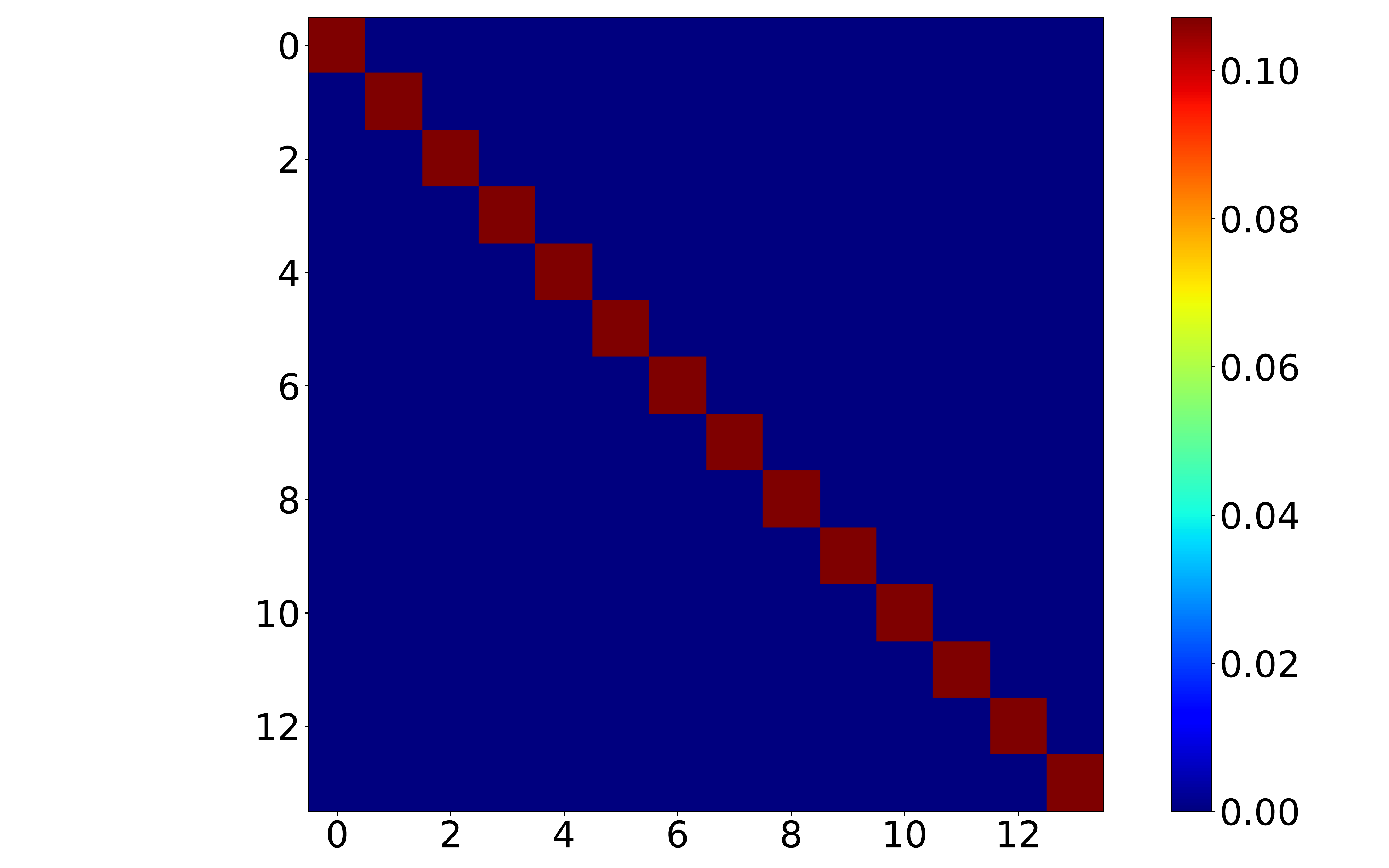}\label{fig:cov_reldiff_highdark}}           

\caption{Group difference covariance matrices in the case of the white (panels a and d) and correlated readout (panels b and e) noise approximations.  In panels (c) and (f) we also present the relative differences. The top (a, b and c) and bottom (c, d and e) panels correspond to an incident flux of $0.001 \ \rm{e}^{-}\rm{s}^{-1}$ (dark conditions) and of $2 \ \rm{e}^{-} \rm{s}^{-1}$ (nominal sky background), respectively. See main text for details.}
\label{fig:cov_matr}
\end{figure*}

\clearpage

\section{Results and discussions}\label{sec:discussions_bias}

\subsection{Group and  group difference covariance matrices}

\subsubsection*{Verification via simulations}\label{sec:simulation}

\indent\indent In order to validate our analytical expressions for the group and group difference covariance matrices in the case of correlated readout noise we have performed Monte Carlo simulations. We have generated a large number of realizations of fake NISP readout noise using the $(1/f)^{\, \alpha}$-like model discussed in Sect.~\ref{sec:fit}. The correlated readout noise simulations are obtained via three steps: 1) we produce realizations of Gaussian white noise in real space, 2) we take the Fourier transform of those and multiply each Fourier component by the square root of the value of the power spectrum model at the same frequency, and 3) we compute the inverse Fourier transform of the modified Fourier components of the readout noise simulation.
From these simulations of readout noise we have constructed fake NISP ramps by adding a cumulative flux contribution as well as the corresponding photon noise assuming a Poisson distribution. We consider here the $\rm{MACC}(15, 16, 11)$ readout mode as chosen for the spectroscopic in-flight operations, for which the effect of the correlated noise is expected to be larger.
As an example, we present in Fig.~\ref{fig:cov_matr_comparison} the group difference covariance matrix as obtained from Eqs.~ (\ref{eq:dkk_corr}) and (\ref{eq:dkl_corr}) (left panel), and the relative difference with respect to the one obtained from Monte Carlo simulations (right panel). The estimates of the covariance matrix were computed for the values of $\sigma$,  $f_{\rm{knee}}$ and $\alpha$ found in Sect.~\ref{sec:fit} for the NISP detector data. The incident flux is set to $1 \ \rm{e}^{-}\rm{s}^{-1}$. We observe very good agreement between the two estimates. We have repeated this comparison for various values of the parameters $\sigma$ , $f_{\rm{knee}}$ and $\alpha$, and for different input fluxes, and obtained the same results. We therefore validate our analytical expressions.

\subsubsection*{White and correlated readout noise covariance matrices}

In this paper, we are interested in studying how using a white noise approximation in the case of a correlated readout noise can impact the on-board estimation of the total flux measured by the \Euclid\ detectors. Therefore, it is interesting to compare the covariance matrix one would obtain for the same correlated input noise in the white and correlated readout noise approximations discussed in Sect.~\ref{sec:corr_equations}. 
The correlated readout noise is obtained via Monte Carlo simulations. We generate mock timelines using the $(1/f)^{\, \alpha}$-like model discussed above with the set of averaged best-fit parameters presented in Sect.~\ref{sec:fit}. The covariance matrix for the correlated noise approximation is computed as described in Sect.~\ref{sec:corr_equations}. For the white noise approximation we start by computing the effective root mean square (rms) of the readout noise. In practice we deduce it from the CDS noise estimated from the simulated timelines of correlated readout noise and impose $\sigma_{\rm{white}} = \rm{CDS}/ \sqrt{2}$. The covariance matrix in the white noise approximation is computed following \citet{2016PASP..128j4504K}. In terms of the signal contribution we have considered two cases: 1) dark conditions and 2) sky background on nominal \Euclid\ flight operations. For the dark conditions we assume an incident flux of $10^{-3} \ \rm{e}^{-} \rm{s}^{-1}$, while for the sky background we consider  $2 \  \rm{e}^{-}\rm{s}^{-1}$, which is within th 10 \% of the expected in-flight background in the NISP \textit{Y}, \textit{J} and \textit{H} filters including zodiacal light and scattered light.  

The main results of this analysis are presented in Fig.~\ref{fig:cov_matr}, where we represent the group difference covariance matrix for the white (left) and correlated (middle) readout noise approximations, and the relative difference between the two (right). In the top and bottom panels of the figure we show the covariance matrices for the dark conditions and nominal sky background cases, respectively.
For all the covariance matrices displayed we observe strong correlation between adjacent group differences as one would expect for non-destructive exposures. For the dark conditions we observe important differences between the white and correlated readout noise approximations. This is due to the fact that the readout noise dominates with respect to the photon noise. We mainly find that the difference between the diagonal and adjacent terms is increased in the correlated readout noise approximation with respect to the white noise one. Furthermore, other off-diagonal terms are not strictly zero for the correlated approximation by contrast to the white noise one. These differences would increase in the case of either steeper power spectrum or a larger $f_{\rm{knee}}$ frequency.
For the nominal background conditions we expect the photon noise contribution to be dominant and therefore we find that the differences between the white and correlated readout noise approximations are smaller. As before, these differences would depend very much on the slope of the readout noise power spectrum and on its $f_{\rm{knee}}$ frequency. Therefore, we think that an accurate characterisation of these parameters will be needed during ground calibration and in-flight commissioning and performance verification phases.

\subsection{On-board flux estimation}\label{sec:bias}

\begin{figure}[t!] 
    \centering 
        \includegraphics[width = 1.\textwidth]{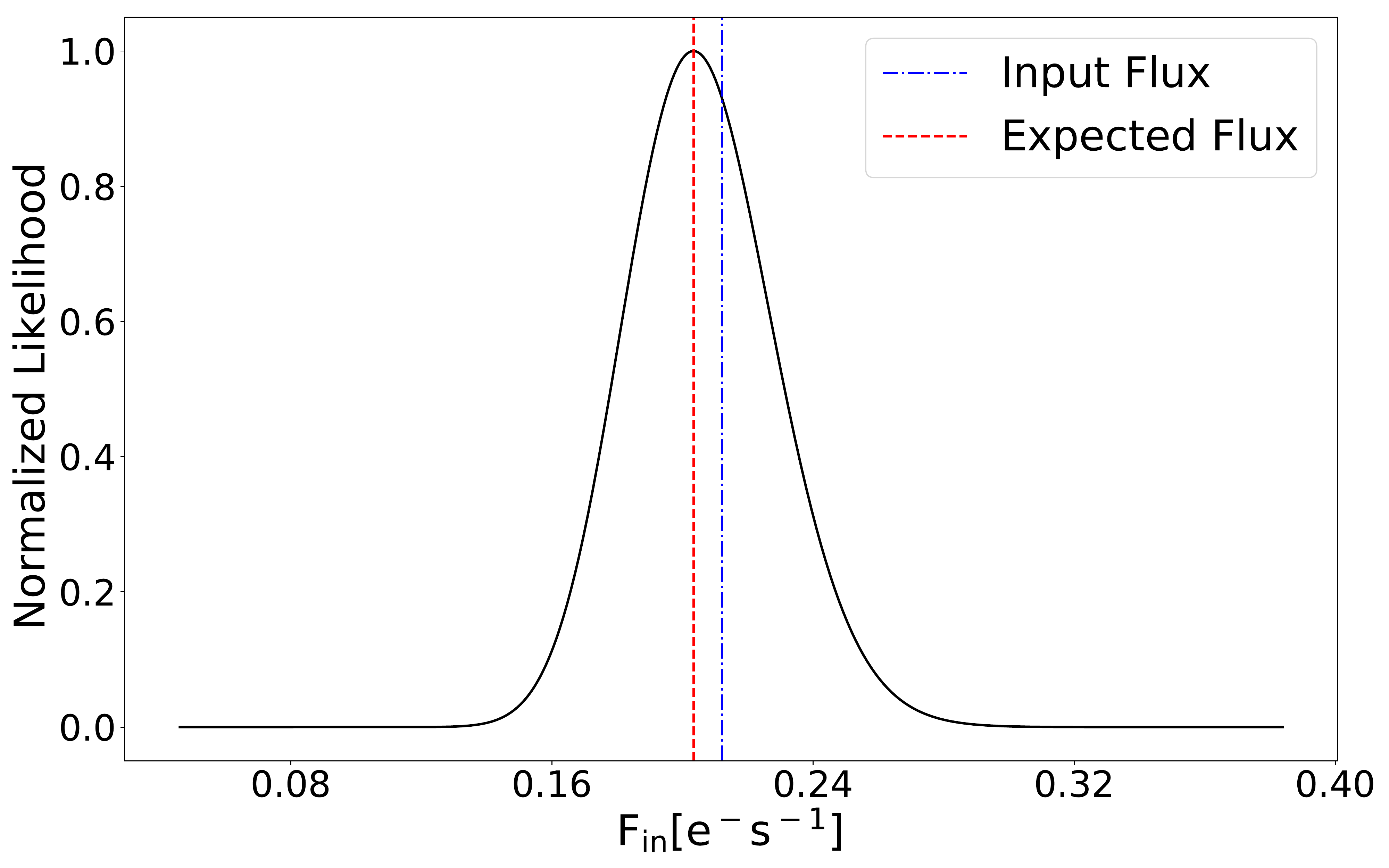} 
        \caption{Example of reconstructed normalized likelihood function (black curve) for a simulation with input sky flux of 0.21 $\rm{e}^{-}\rm{s}^{-1}$ (blue dashed vertical line)  and in the case of correlated readout noise. The best-fit flux is shown as a vertical red dashed line. 
\label{fig:likelihood}}
\end{figure} 

\begin{figure}[h!] 
    \centering 
        \includegraphics[width = 1.\textwidth]{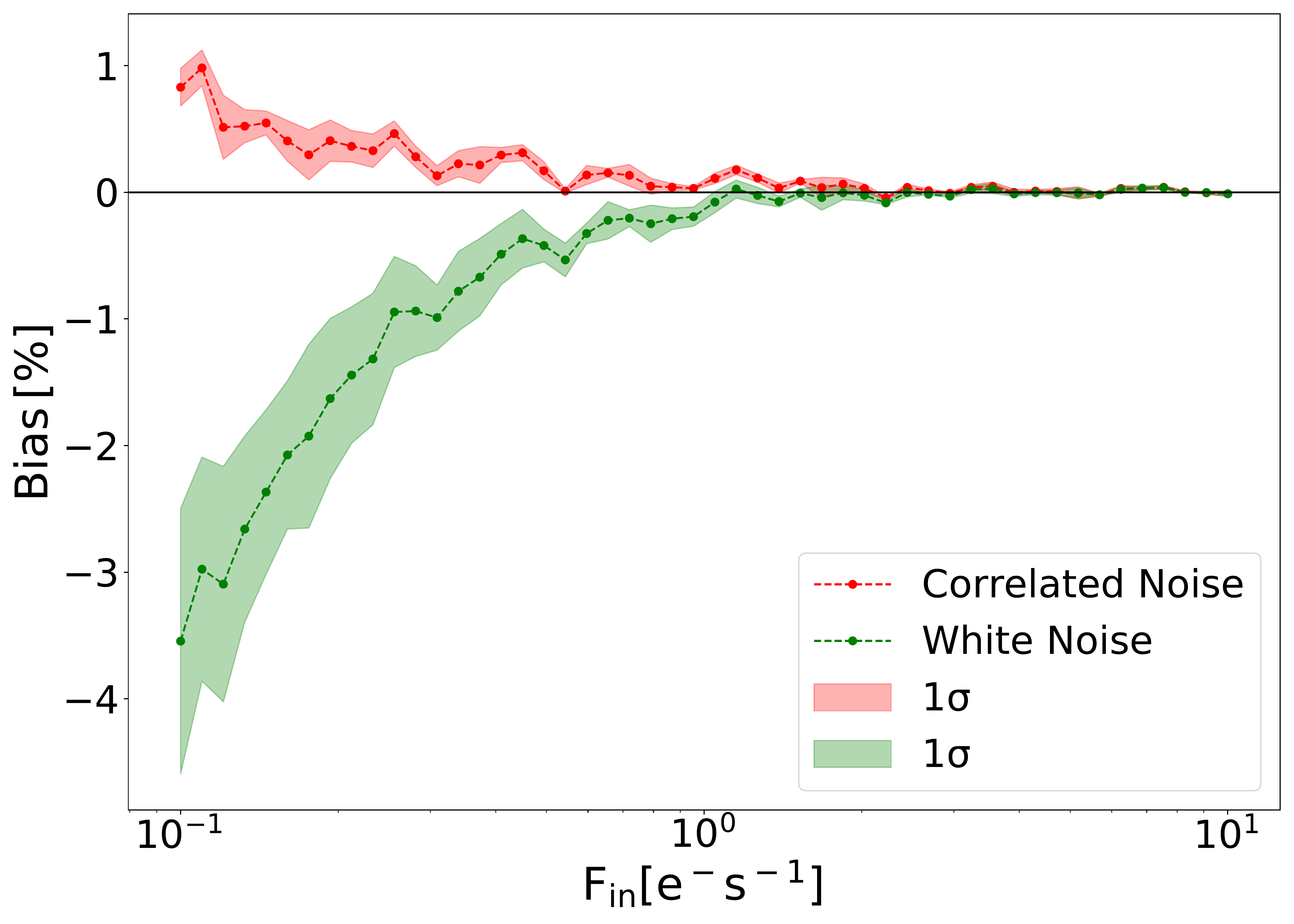} 

        \caption{Expected bias of the flux estimator assuming $\rm{MACC}(15, 16, 11)$ and the median noise parameters: $\sigma = 19.70$ $\rm{e}^{-}\rm{Hz}^{-0.5}$, $f_{\rm{knee}} = 5.2 \times 10^{-3} \ \rm{Hz}$ and $\alpha = 1.24$.}
\label{fig:bias}

\end{figure}

As discussed above during \Euclid\ flight operations the white noise approximation will be used to estimate the sky flux in each of the NISP array pixels and only the sky image will be transferred to Earth. In the presence of correlated noise we expect the estimate of the flux to be biased. To evaluate this bias we have constructed mock simulations of the sky emission by assuming a constant sky signal and adding realistic realisations of the readout noise in the NISP detectors. For the latter we use the $(1/f)^{\, \alpha}$-like model and best-fit parameters discussed in Sect.~\ref{sec:fit}, and the simulation procedure described above.
We explore the sky signal in the range 0.1--100 $\rm{e}^{-}\rm{s}^{-1}$, i.e., from low dark values to bright objects.

For each value of the sky signal we construct $10 \,000$ mock ramps. For each of the simulated ramps we estimate the sky flux from the group differences using: 1) the white noise approximation flux estimator (WNA, hereafter) developed by \citep{2016PASP..128j4504K}, and 2) the maximum likelihood approach discussed in Eq.~\ref{eq:likelihood} assuming the noise group difference covariance matrix presented in Sect.~\ref{sec:corr_equations}, (CNA, hereafter).
In practice, the maximum likelihood flux estimate for 1) is obtained using equation 11 in \citet{2016PASP..128j4504K}.
For the CNA case the maximum likelihood flux estimate is obtained using a simple grid approach. For illustration, an example of the reconstructed likelihood function for a sky flux of 0.21 $\rm{e}^{-}\rm{s}^{-1}$ is presented as solid black curve in Fig.~\ref{fig:likelihood}. The best-fit value and input value are indicated as vertical dashed red and dashed blue lines, respectively. We observe that even when using the expected group difference covariance matrix for the correlated readout noise there is still a small bias in the estimate of the sky flux. 

In Fig.~\ref{fig:bias} we show the relative bias, $\frac{\rm{F}_{\rm{out}}-\rm{F}_{\rm{in}}}{\rm{F}_{\rm{in}}}$, in percent for both the CNA (red line and dots) and the WNA (green line and dots) cases as a function of the background flux $\rm{F}_{\rm{in}}$. Uncertainties in the measured bias are given by the filled red (CNA) and green (WNA) areas as computed from the Monte Carlo simulations.
For low background flux values (below $1 \ \rm{e}^{-} \rm{s}^{-1}$) we find that the maximum bias for CNA is under $1\%$ and about four times smaller than the WNA one. For fluxes above $1 \ \rm{e}^{-} \rm{s}^{-1}$ the bias in both cases are equivalent and below 0.1 $\%$. For sky observations we expect a background flux of about 1--2 $\ \rm{e}^{-}\rm{s}^{-1}$, in the region where the bias is expected to be small. However, we have observed using the dispersion over the set of Monte Carlo simulations that the WNA systematically underestimates the uncertainties by a factor ranging from 2 to 5. However, for CNA the dispersion on the simulations and the measured uncertainties are consistent.

\section{Summary and conclusion}\label{sec:conclusion}

The \Euclid\ satellite mission will be a fundamental tool for cosmology and infrared astronomy thanks to its near infrared photometer and spectrometer instrument, NISP, which consists of 16 NIR sensitive H2RG detector arrays of $2048\times2048$ pixels each. The NISP is designed to observe very faint distant galaxies and therefore requires a low and well characterized readout noise. 

In this paper we have studied the readout noise associated with the NISP detectors taking advantage of long exposures (few hours) performed during laboratory dark tests at the CPPM cryogenic facilities. We have found that the NISP readout noise is correlated and can be well characterized by a $(1/f)^{\, \alpha}$-like model with a typical knee frequency of  $f_{\rm{knee}} = \paren{5.2^{+1.8}_{-1.3}} \times 10^{-3} \ \rm{Hz}$ and a low frequency component with slope $\alpha = 1.24 ^{+0.26}_{-0.21}$. From this we conclude that the readout noise of the NISP detectors has non-negligible correlation within the typical in-flight NISP exposure time (575 s).

Infrared instruments, and in particular the NISP, acquire data using the MACC readout mode, which consists of a series of non-destructive exposures averaged into groups that form a ramp. The input flux in the detectors can then be obtained from the slope of the ramp using maximum likehood estimators, which generally assume white readout noise \citep{kubik2015optimization, 2016PASP..128j4504K}. Here, we have extended these estimators to the case of correlated readout noise. Analytical expressions for the group and group difference covariance matrices are presented for the case of $(1/f)^{\, \alpha}$-like correlated readout noise. These have been validated via Monte Carlo simulations. 
 
Finally, we have performed Monte Carlo simulations of the in-flight expected NISP detector signal and noise, including a realistic background signal and correlated readout noise as measured on the ground calibration tests. From these simulations we have been able to estimate the expected bias in the on-board flux estimates during in-flight operations, for which white readout noise is assumed. We find that for low background the flux bias can be up to four times larger than when accounting for the correlation in the readout noise. Nevertheless, this bias is negligible  for typical sky background signals. Therefore, we expect no significant bias in the on-board fluxes measured by \Euclid.

\begin{acknowledgements}
\AckEC
\end{acknowledgements}

\newpage
\bibliography{bib_nisp.bib}
\appendix

\section{Detailed computation of the group noise covariance matrix}
\label{sec:appendix}

We detail here the computation of the group noise covariance matrix forcorrelated readout noise described by \ref{eq:correlationfunc}. We concentrate in the correlated readout noise terms. Other terms can be found in~\cite{kubik2015optimization}.

\noindent The group noise covariance matrix is given by
\begin{equation}
\begin{aligned}
C_{kk}&=(k-1)\,\mathcal{D}+(k-1)(m-1)f+f\frac{(m+1)(2m+1)}{6m}  \\ 
&+\frac{1}{m^2}\sum_{i=1}^{m}\sum_{j=1}^{m}\delta \rho_i^{(k)}\delta \rho_j^{(k)}
\end{aligned}
\end{equation}
\noindent for the diagonal term, and
\begin{equation}
\begin{aligned}
C_{kl} = \langle \delta G_k\delta G_l \rangle & =                                      
(k-1)\,\mathcal{D}+(k-1)(m-1)f+f\frac{(m+1)}{2}  \\
&+
\left<\left[\frac{1}{m}\sum_{i=1}^{m}\delta \rho_i^{(k)}\right] \; \left[\frac{1}{m}\sum_{j=1}^{m}\delta \rho_j^{(l)}\right]\right>
\end{aligned}
\end{equation}

\noindent for the off-diagonal elements. And the readout noise is
\begin{equation}
\begin{aligned}
&\left< \left[\frac{1}{m}\sum_{i=1}^{m}\delta \rho_i^{(k)}\right] \, \left[\frac{1}{m}\sum_{j=1}^{m}\delta \rho_j^{(l)}\right]\right> \\
&=\frac{1}{m^2} \sum_{i=1}^{m}\sum_{j=1}^{m} \lbrace \rho_i^{(k)}\rho_j^{(l)} \rbrace  \\
&=\frac{1}{m^2}\sum_{i=1}^{m} C[|(l-k)(m+d)+(i-i)| \, t_{\rm{frame}}]  \\
&+\frac{1}{m^2}\sum_{i=1}^{m}\sum_{j=1,j\neq i}^{m} C\left(|(l-k)(m+d)+(j-i)| \, t_{\rm{frame}}\right)  \\
&=\frac{1}{m^2} mC\left((l-k)(m+d) \, t_{\rm{frame}}\right)  \\
&+\frac{1}{m^2}\sum_{i=1}^{m}\sum_{j=1,j\neq i}^{m} C\left(|(l-k)(m+d)+(j-i)| \, t_{\rm{frame}}\right) \\
&=\frac{1}{m} C\left((l-k)(m+d) \, t_{\rm{frame}}\right)   \\
&+\frac{1}{m^2} \sum_{i=1}^{m}\sum_{j=1,j\neq i}^{m}C\left(|(l-k) \, (m+d)+ |(j-i)| \, t_{\rm{frame}}\right).
\end{aligned}
\end{equation}

\noindent For the diagonal terms, $l = k$ , the readout noise term is
\begin{equation}
\begin{aligned}
&\left< \left[\frac{1}{m}\sum_{i=1}^{m}\delta \rho_i^{(k)}\right] \; \left[\frac{1}{m}\sum_{j=1}^{m}\delta \rho_j^{(k)}\right]\right>  \\
&= \frac{1}{m^2} \left[mC(0)  + \sum_{i=1}^{m}\sum_{j=1,j\neq i}^{m}C\left(|j-i| \, t_{\rm{frame}}\right) \right]  \\
&= \frac{1}{m^2} \left[ mC(0)+2\sum_{i=1}^{m-1}(m-i)C\left(i \, t_{\rm{frame}}\right) \right]
\end{aligned}
\end{equation}

\noindent and then
\begin{equation}
\begin{aligned}
C_{kk} &=(k-1)\,\mathcal{D}+(k-1)(m-1)f+f\frac{(m+1)(2m+1)}{6m}  \\ 
&+\frac{1}{m^2}\left[mC(0)+2\sum_{i=1}^{m-1}(m-i)C(i \, t_{\rm{frame}})\right].
\end{aligned}
\end{equation}

\noindent For the off-diagonal terms, $l \neq k$ , we have

\begin{equation}
\begin{aligned}
&\left<\left[\frac{1}{m}\sum_{i=1}^{m}\delta \rho_i^{(k)}\right] \; \left[\frac{1}{m}\sum_{j=1}^{m}\delta \rho_j^{(l)}\right]\right> \nonumber \\
&= \frac{1}{m^2} mC\left((l-k)(m+d) \, t_{\rm{frame}}\right) \nonumber\\
&+\frac{1}{m^2} \sum_{i=1}^{m}\sum_{j=1,j\neq i}^{m}C\left(|(l-k)(m+d) \, t_{\rm{frame}}+(j-i)| \, t_{\rm{frame}}\right)
\end{aligned}
\end{equation}

\noindent and then
\begin{equation}
\begin{aligned}
C_{kl} &=  \langle \delta G_k\delta G_l \rangle =                                      
(k-1)\,\mathcal{D}+(k-1)(m-1)f+f\frac{(m+1)}{2}  \\
&+\frac{1}{m} C\left((l-k)(m+d) \, t_{\rm{frame}} \right) \\
&+\frac{1}{m^2} \sum_{i=1}^{m}\sum_{j=1,j\neq i}^{m}C\left(|(l-k)(m+d) \, t_{\rm{frame}}+(j-i)| \, t_{\rm{frame}}\right).
\end{aligned}
\end{equation}

\subsection{Group differences}
\subsubsection{Diagonal Terms}
We can write the diagonal terms of the group differences covariance as
\begin{equation}
\begin{aligned}
D_{kk}&= d + f\frac{(m-1)(2m-1)}{3m}  \\
&+\frac{1}{m^2}\sum_{i=1}^{m} \sum_{j=1}^{m} \langle\delta \rho_i^{(k+1)}\delta \rho_j^{(k+1)}\rangle \\
&+ \frac{1}{m^2}\sum_{i=1}^{m}\sum_{j=1}^{m}\angle\delta \rho_i^{(k)}\delta \rho_j^{(k)}\rangle  \\
&- \frac{1}{m^2}\sum_{i=1}^{m}\sum_{j=1}^{m}\langle\delta \rho_i^{(k+1)}\delta \rho_j^{(k)}\rangle \\
&- \frac{1}{m^2}\sum_{i=1}^{m}\sum_{j=1}^{m}\langle\delta \rho_i^{(k)}\delta \rho_j^{(k+1)}\rangle.
\end{aligned}
\end{equation}

\noindent We compute now each term of the correlated readout noise contribution:

\begin{equation}
\begin{aligned}
&\sum_{i=1}^{m}\sum_{j=1}^{m}\langle\delta \rho_i^{(k+1)}\delta \rho_j^{(k+1)}\rangle  \\
&= \sum_{i=1}^{m}\sum_{j=1}^{m}\langle\delta \rho_i^{(k)}\delta \rho_j^{(k)}\rangle \\
&=\sum_{i=1}^{m}\sum_{j=1}^{m} C\left(|j-i| \, t_{\rm{frame}}\right);
\end{aligned}
\end{equation}

\begin{equation}
\begin{aligned}
 &\sum_{i=1}^{m}\sum_{j=1}^{m} \langle \delta \rho_i^{(k+1)}\delta \rho_j^{(k)}\rangle  \\
 &= \sum_{i=1}^{m}\sum_{j=1}^{m} C\left(|m+d+j-i| \, t_{\rm{frame}}\right)  \\
 &= \sum_{i=1}^{m}\sum_{j=1}^{m} C\left(|j-i-m-d)| \, t_{\rm{frame}}\right);
\end{aligned}
\end{equation}

\begin{equation}
\begin{aligned}
 &\sum_{i=1}^{m}\sum_{j=1}^{m}\langle\delta \rho_i^{(k)}\delta \rho_j^{(k+1)}\rangle  \\
 &= \sum_{i=1}^{m}\sum_{j=1}^{m} C\left(|m+d+j-i| \, t_{\rm{frame}}\right) \\
 &= \sum_{i=1}^{m}\sum_{j=1}^{m} C\left(|j - i + m+d| \, t_{\rm{frame}}\right). 
\end{aligned}
\end{equation}

\noindent Finally, the group difference diagonal covariance matrix is
\begin{equation}
\begin{aligned}
D_{kk}&=
\,\mathcal{D} + f\frac{(m-1)(2m-1)}{3m}  \\
&+2 \frac{1}{m^2}\sum_{i=1}^{m}\sum_{j=1}^{m} C\left(|j-i| \, t_{\rm{frame}}\right)  \\
&-\frac{1}{m^2}\sum_{i=1}^{m}\sum_{j=1}^{m} C\left(|j - i + m+d| \, t_{\rm{frame}}\right)  \\
&- \frac{1}{m^2}\sum_{i=1}^{m}\sum_{j=1}^{m} C\left(|j - i - m-d| \, t_{\rm{frame}}\right).
\end{aligned}
\end{equation}

\subsubsection{Off-diagonal terms}

The off-diagonal terms are given by
\begin{equation}
\begin{aligned}
D_{kl}&=\frac{f}{6m}(m^2 -1)\delta_{(k+1)l}  \\
&+\frac{1}{m^2}\sum_{i=1}^{m}\sum_{j=1}^{m}
\langle\delta \rho_i^{(k+1)}\rho_j^{(l+1)}\rangle \\
&+\frac{1}{m^2}\sum_{i=1}^{m}\sum_{j=1}^{m}
\langle\delta \rho_i^{(k)}\rho_j^{(l)}\rangle   \\
&-\frac{1}{m^2}\sum_{i=1}^{m}\sum_{j=1}^{m}
\langle\delta \rho_i^{(k+1)}\rho_j^{(l)}\rangle \\
&-\frac{1}{m^2}\sum_{i=1}^{m}\sum_{j=1}^{m}
\langle\delta \rho_i^{(k)}\rho_j^{(l+1)}\rangle.
\end{aligned}
\end{equation}

\noindent We compute now each term of the correlated readout noise contribution:

\begin{equation}
\begin{aligned}
&\sum_{i=1}^{m}\sum_{j=1}^{m}
\langle\delta \rho_i^{(k+1)}\rho_j^{(l+1)}\rangle \\
&=\sum_{i=1}^{m}\sum_{j=1}^{m}
\langle\delta \rho_i^{(k)}\rho_j^{(l)}\rangle  \\
& =\sum_{i=1}^{m}\sum_{j=1}^{m} C\left(|(l-k)(m+d)+(j-i)| \ ; t_{\rm{frame}}\right);
\end{aligned}
\end{equation}

\begin{equation}
\begin{aligned}
&\sum_{i=1}^{m}\sum_{j=1}^{m}
\langle\delta \rho_i^{(k+1)}\rho_j^{(l)}\rangle  \\
& = \sum_{i=1}^{m}\sum_{j=1}^{m} C\left(|(l-k-1)(m+d)+(j-i)| \, t_{\rm{frame}}\right);
\end{aligned}
\end{equation}
 
\begin{equation}
\begin{aligned}
&\sum_{i=1}^{m}\sum_{j=1}^{m}
\langle\delta \rho_i^{(k)}\rho_j^{(l+1)}\rangle \\
& = \sum_{i=1}^{m}\sum_{j=1}^{m} C\left(|(l-k+1)(m+d)+(j-i)| \, t_{\rm{frame}}\right).
\end{aligned}
\end{equation}

\noindent Finally, the group difference off-diagonal covariance matrix is
\begin{equation}
\begin{aligned}
D_{kl}&=\frac{f}{6m}(m^2 -1)\delta_{(k+1)l}  \\
&+2\frac{1}{m^2}\sum_{i=1}^{m}\sum_{j=1}^{m} C\left(|(l-k)(m+d)+(j-i)| \, t_{\rm{frame}}\right)\\
&-\frac{1}{m^2}\sum_{i=1}^{m}\sum_{j=1}^{m} C\left(|(l-k-1)(m+d)+(j-i)| \, t_{\rm{frame}}\right)  \\
&-\frac{1}{m^2}\sum_{i=1}^{m}\sum_{j=1}^{m} C\left(|(l-k+1)(m+d)+(j-i)| \, t_{\rm{frame}}\right).
\end{aligned}
\end{equation}

\end{document}